\documentclass[twocolumn,aps,superscriptaddress]{revtex4-1}
\usepackage{graphicx,amsmath,amsfonts,bbold,dsfont,bbm}
\begin{document}

\title{Fingerprints of spin-current physics on magnetoelectric response \\ in the spin-$1/2$ magnet Ba$_2$CuGe$_2$O$_7$}

\author{Ryota Ono}
\affiliation{Graduate School of Science and Engineering, Chiba University, 1-33 Yayoi-cho, Inage-ku, Chiba-shi 265-8522, Japan}
\author{Sergey Nikolaev}
\affiliation{Institute of Innovative Research, Tokyo Institute of Technology, 4259 Nagatsuta, Midori, Yokohama 226-8503, Japan}
\affiliation{International Center for Materials Nanoarchitectonics,
National Institute for Materials Science, 1-1 Namiki, Tsukuba,
Ibaraki 305-0044, Japan}
\author{Igor Solovyev}
\email{SOLOVYEV.Igor@nims.go.jp}
\affiliation{International Center for Materials Nanoarchitectonics,
National Institute for Materials Science, 1-1 Namiki, Tsukuba,
Ibaraki 305-0044, Japan}
\affiliation{Department of Theoretical Physics and Applied Mathematics, Ural Federal University,
Mira str. 19, 620002 Ekaterinburg, Russia}
\affiliation{Institute of Metal Physics, S. Kovalevskaya str. 18, 620108 Ekaterinburg, Russia}

\date{\today}

\date{\today}
\begin{abstract}
As is well known, the single-site anisotropy vanishes in the spin-$1/2$ compounds as a consequence of fundamental Kramers degeneracy. We argue, rather generally, that similar property holds for the magnetically induced electric polarization $\boldsymbol{P}$, which should depend only on the relative orientation of spins in the bonds but not on the direction of each individual spin. Thus, for insulating multiferroic compounds, $\boldsymbol{P}$ can be decomposed in terms of pairwise isotropic, antisymmetric, and anisotropic symmetric contributions, which  can be rigorously derived in the framework of the superexchange (SE) theory, in an analogy with the spin Hamiltonian. The SE theory also allows us to identify the microscopic mechanism, which stands behind each contribution. The most controversial and intriguing one -- concerning the form, appearances, and implications to the properties of real compounds -- is antisymmetric or spin-current mechanism. In this work, we propose that, within the SE theory, the disputed magnetoelectric (ME) properties of tetragonal Ba$_2$CuGe$_2$O$_7$, representing the lattice of magnetic Cu$^{2+}$ ions in the tetrahedral environment, can be explained solely by the spin-current mechanism, while other contributions are either small or forbidden by symmetry. First, after analysis of the symmetry properties of the SE Hamiltonian and corresponding parameters of electric polarization, we explicitly show how the cycloidal spin order induces the experimentally observed electric polarization in the direction perpendicular to the tetragonal plane, which can be naturally explained by the spin-current mechanism operating in the out-of-plane bonds. Then, we unveil previously overlooked ME effect, where the application of the magnetic field perpendicular to the plane not only causes the incommensurate-commensurate transition, but also flips the electric polarization into the plane due to the spin-current mechanism operating in the neighboring bonds within this plane. In both cases, the magnitude and direction of $\boldsymbol{P}$ can be controlled by rotating the spin pattern in the tetragonal plane. Our analysis is based on a realistic spin model, which was rigorously derived from the first-principles electronic structure calculations and supplemented with the new algorithm for the construction of localized Wannier functions obeying the crystallographic symmetry of Ba$_2$CuGe$_2$O$_7$.
\end{abstract}

\maketitle

\section{\label{sec:Intro} Introduction}
\par The magnetoelectric (ME) effect, namely the ability of certain antiferromagnetic (AFM) substances obeying certain symmetry properties to become ferroelectric upon applying the magnetic field and ferromagnetic upon applying the electric field~\cite{DzyaloshinskiiME}, is the key fundamental phenomenon opening a route for the creation of new-type electronic devices utilizing such cross-control as the basic principle of their functionality~\cite{EerensteinNature,TokuraScience,KimuraARMR}. The broad interest in this topic has revived again in early 2000s, after the discovery of multiferroics~\cite{Kimura_TbMnO3}: the materials in which the ferroelectricity coexists with a long-range magnetic order without electric or magnetic field and in many cases is driven by this order~\cite{CheongMostovoy,Khomskii}. Hence, the magnetic order should break the inversion symmetry. The simplest spin pattern satisfying this requirement is the spin spiral~\cite{Sandratskii}. Although this choice is not unique, the materials potentially possessing spin-spiral textures have attracted a great deal of attention in a bid to search for new multiferroics~\cite{Kimura_TbMnO3,TokuraSeki}.

\par The microscopic understanding of the origins and driving forces responsible for the ME coupling is vitally important as it should serve as a guide for the analysis and predictions of new such materials and phenomena. Then, what do we know about the dependence of electric polarization on the magnetization? Surprisingly still not much in comparison with the progress achieved along the same line for the description of energy in terms of magnetic interactions, for which there is a long-established Heisenberg model~\cite{Heisenberg,Anderson,JHeisenberg}, which can be further refined by including the spin-orbit (SO) interaction related terms, such as the single-site anisotropy, the antisymmetric Dzyaloshinskii-Moriya (DM) exchange~\cite{Dzyaloshinskii_weakF,Moriya_weakF}, and the bond-dependent symmetric anisotropic exchange~\cite{Khaliullin2009}. The symmetry principles for all these magnetic interactions are well documented, including nonexistence of the single-site anisotropy for the spin $1/2$: one of basic properties of spin systems resulting from the fundamental Kramers' degeneracy. For localized spins in insulating materials such model can be rigorously formulated in the framework of superexchange (SE) theory~\cite{Anderson}: the technique, which is also well established today~\cite{Khaliullin2009,KugelKhomskii,PRB2015b}.

\par Can the same strategy be applied for the description of electric polarization $\boldsymbol{P}$? Indeed, already in 1968, Moriya, on the basis of symmetry considerations, has proposed a spin model for the polarization, which had all main ingredients including the single-site, isotropic, antisymmetric, and symmetric anisotropic ones, in full analogy with the magnetic energy~\cite{Moriya1968}. However, the rigorous microscopic theories behind this model were missing, so that the behavior of electric polarization in multiferroics was typically discussed~\cite{TokuraSekiNagaosa} in terms of separate phenomenological rules expected for the exchange striction (of either symmetric~\cite{Ca3CoMnO6} or antisymmetric~\cite{SergienkoPRB} type), the spin current in spiral magnets~\cite{KNB,Mostovoy}, or the metal-ligand hybridization change~\cite{CuFeO2_Arima}. The `spin current' in this context means the noncollinear alignment of spins, which can be related to the spin flows in the direction perpendicular to the spins~\cite{BrunoDugaev}. Therefore, many properties of noncollinear magnets can be related to such spin current~\cite{Kikuchi}. Although each theory of magnetically induced ferroelectricity has certain logic behind, the situation remains very controversial as many experimental data can be interpreted from completely different standpoints, involving different scenaria of the ME coupling~\cite{MurakawaPRL,MurakawaPRB,Cu2OSeO3,Seki2012,YWLee,JTZhang,YNii,PRB2014}. Furthermore, there is a growing understanding that the phenomenological spin-curent theory of the ME coupling needs to be revised as in the present form it fails to capture many important phenomena, which have been discovered afterwards, such as the multiferroicity in the proper-screw spiral magnets~\cite{MnI2_Xiang,PRB2017}.

\par The story of Ba$_2$CuGe$_2$O$_7$ presents a typical example of this controversy. It is a canonical spiral magnet, crystallizing in the noncentrosymmetric but nonpolar $P\overline{4}2_{1}m$ structure, where the chiral magnetic order is driven by DM interactions~\cite{Zheludev1996}. The material exhibits a number of interesting phenomena originating from the interplay of the DM interactions and the exchange anisotropy in the external magnetic field and resulting in a complex phase diagram~\cite{Zheludev1997,Zheludev1998,Zheludev1999,Zheludev2012,Chovan}. Although ferroelectricity is not allowed by the crystallographic symmetry, it can be induced by the cycloidal spin order, so that Ba$_2$CuGe$_2$O$_7$ can potentially become multiferroic. Such multiferroicity was indeed observed in 2009 by Murakawa~\textit{et al.}~\cite{MurakawaPRL}. Furthermore, these authors have clearly demonstrated how the ferroelectric single domains can be generated by an application of magnetic field. Alternatively, the magnetic domains with the given spin-spiral propagation vector $\boldsymbol{q}$ can be switched by the electric field. Although the measured effect was small, it is of great fundamental importance.

\par The discovery of magnetically induced ferroelectric activity in Ba$_2$CuGe$_2$O$_7$ was spurred by general search for multiferroics with spiral magnetic texture~\cite{Kimura_TbMnO3,TokuraSeki}, which was believed to be primarily responsible for this effect following suggestions of the phenomenological spin-current theories~\cite{KNB,Mostovoy}. Ba$_2$CuGe$_2$O$_7$ was certainly a potential candidate in this search. However, soon after discovery of multiferroicity in Ba$_2$CuGe$_2$O$_7$, somewhat similar behavior was found in its sister materials, Ba$_2$CoGe$_2$O$_7$~\cite{MurakawaPRL2}. Unlike Ba$_2$CuGe$_2$O$_7$, Ba$_2$CoGe$_2$O$_7$ forms a commensurate AFM spin texture with no sign of the spin-spiral order. Nevertheless, the electric polarization observed in Ba$_2$CoGe$_2$O$_7$ was at least an order of magnitude larger than in Ba$_2$CuGe$_2$O$_7$. This has led to the conclusion that the origin of the ferroelectric activity, both in Ba$_2$CoGe$_2$O$_7$ and in Ba$_2$CuGe$_2$O$_7$, is not related to the chiral order, but caused by another mechanism of the spin-dependent metal-ligand hybridization~\cite{MurakawaPRB}, which is basically a single-site property as it depends on individual directions of the localized spins but not on the correlations between the spins.

\par Since the ferroelectricity is the property of insulating substances~\cite{footnote4}, we consider that it is natural to extend the SE theory in order to deal, besides the exchange interactions, also with the magnetic dependencies of $\boldsymbol{P}$~\cite{PRB2017,PRB2019}, and start for these purposes with the general theory of electric polarization in periodic systems~\cite{FE_theory1,FE_theory2,FE_theory3}. In this work, we elaborate this strategy for Ba$_2$CuGe$_2$O$_7$ and argue that it can indeed resolve many controversial issues of Ba$_2$CuGe$_2$O$_7$ and other materials with the chiral magnetic order. First, by extending the analysis for the magnetocrystalline anisotropy energy, we show that there should be no single-site contribution to the electric polarization for the spin $1/2$. This simple but fundamental principle basically excludes the spin-dependent metal-ligand hybridization scenario from the analysis of ferroelectric activity of magnetic compounds built from the spin-$1/2$ ions, such as Cu$^{2+}$, Ni$^{3+}$, V$^{4+}$, and Ti$^{3+}$. Then, we argue that the antisymmetric spin-current mechanism (provided that it is properly defined~\cite{PRB2017}) is almost solely responsible for the ME properties in Ba$_2$CuGe$_2$O$_7$, while other contributions (for instance, due to the isotropic coupling) are either small or forbidden by symmetry. Thus, Ba$_2$CuGe$_2$O$_7$ provides a unique platform for realization and exploration of the ME effects arising solely from the spin-current mechanism. Finally, we predict a new ME effect in Ba$_2$CuGe$_2$O$_7$, where the application of the magnetic field along the crystallographic $z$ axis not only causes the incommensurate-commensurate transition, but also flips the polarization from the $z$ axis into the tetragonal $xy$ plane.

\par It is worth mentioning that the interest in Ba$_2$CuGe$_2$O$_7$ is not limited to its multiferroic properties. Another interesting aspect of Ba$_2$CuGe$_2$O$_7$ is the coexistence of chiral magnetic structures and weak ferromagnetism, which are driven by two types of DM interactions existing in the systems with the $P\overline{4}2_{1}m$ symmetry~\cite{Bogdanov}. In addition to the regular spin-spiral state, this symmetry allows the formation of \textit{anti}skyrmion spin textures, which can play an important role in future spintronic applications~\cite{Nayak,Huang}.

\par Furthermore, the lattice of the Cu$^{2+}$ ions in the tetrahedral environment provides an interesting possibility for the realization of the SO Mott state in $3d$ oxides~\cite{PRB2018}. Typically, such state is regarded to be a prerogative of heavy-elements compounds with strong SO interaction. However, for the one-hole systems composed of the Cu$^{2+}$ ions, the 1st and 2nd Hund's rules are no longer applicable and the spin-orbital character of a single hole is fully specified by the SO interaction competing with the crystal field (which is comparatively weak in the tetrahedral environment). It gives us a possibility to speak of such systems as $3d$ analogues of $5d$ iridates, which have attracted a great deal of attention~\cite{Khaliullin2009,Kim2008}. Such behavior was recently predicted in CuAl$_2$O$_4$~\cite{PRB2018}, and Ba$_2$CuGe$_2$O$_7$ is another interesting candidate along this line.

\par The rest of the paper is organized as follows. In Sec.~\ref{sec:GGA} we briefly discuss the details of the electronic structure in the generalized gradient approximation (GGA), which is used as the starting point for the construction of electronic and then spin models of Ba$_2$CuGe$_2$O$_7$. A special attention is paid to calculations of Wannier functions obeying the correct crystallographic symmetry. The Wannier functions play a very important role in the construction of the spin model, because both the magnetic energy and the electric polarization in the SE theory are formulated in terms of these functions, where maintaining the correct crystallographic symmetry is one of the crucial factors. The commonly used maximally localized Wannier function (MLWF) technique~\cite{WannierRevModPhys,wannier90} breaks this symmetry and we found the situation to be especially ill-behaved for non-centrosymmetric compounds like Ba$_2$CuGe$_2$O$_7$. Instead, we propose a simple, but very efficient refinement of the MLWF method, which allows us to properly tackle the symmetry issue. Then, in Sec.~\ref{sec:symmetry}, we discuss the spin model for the exchange energy and the electric polarization. Particularly, in Sec.~\ref{sec:general} we explain how both models can be formulated and constructed in terms of the Wannier functions; in Sec.~\ref{sec:polsi} we prove non-existence of single-site contributions for the spin $1/2$; and in Secs.~\ref{sec:ex} and \ref{sec:pol} we discuss symmetry properties of the parameters of exchange interactions and electric polarization, respectively. In Sec.~\ref{sec:model} we consider the exchange interactions as obtained in the 1-orbital and more general 5-orbital models and their relevance to the magnetic structure and properties of Ba$_2$CuGe$_2$O$_7$. Although the 1-orbital model already captures the behavior of the exchange interactions, the orbital degrees of freedom are essential for the analysis of $\boldsymbol{P}$, which is considered in Sec.~\ref{sec:elpol}. Particularly, Sec.~\ref{sec:sppol} deals with the behavior of electric polarization induced by the cycloidal spin order, while the reorientation of polarization associated by the incommensurate-commensurate transition in the magnetic field~\cite{Zheludev1997,Zheludev1998} is considered in Sec.~\ref{sec:meeffect}. We unveil the microscopic origin of such magnetic state dependence of the polarization and show that in both cases it is caused by the spin-current mechanism. In Sec.~\ref{sec:other}, we will present some critical analysis by considering other mechanisms and contributions to $\boldsymbol{P}$ in connection with the phenomenological theories~\cite{SergienkoPRB,KNB,Mostovoy}. Finally, in Sec.~\ref{sec:summary}, we summarize our work.

\begin{center}
\section{\label{sec:GGA} Electronic band structure and model Hamiltonians}
\end{center}
\par
\begin{figure}[b]
\begin{center}
\includegraphics[width=0.40\textwidth]{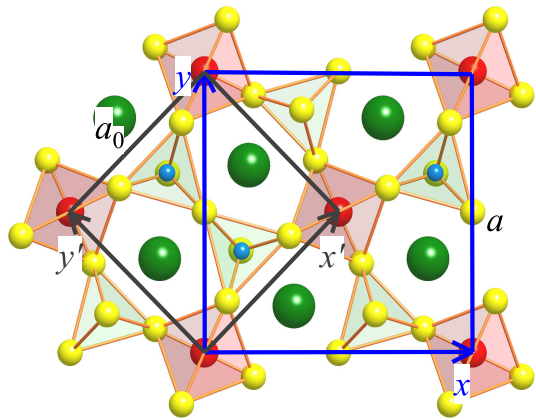}
\end{center}
\caption{
Crystal structure of Ba$_2$CuGe$_2$O$_7$ in the tetragonal plane. The Cu and O atoms are indicated by the medium red and yellow spheres, respectively, the Ba atoms are indicated by the big green spheres, and the Ge atoms are indicated by the small blue spheres. The CuO$_4$ and GeO$_4$ tetrahedra are colored red and green, respectively. The regular unit cell ($xy$, with the lattice parameter $a$) containing two formula units, is shown by blue line. The smaller unit cell ($x'y'$, with the lattice parameter $a_0 = \frac{1}{\sqrt{2}}a$), containing one formula unit and describing the periodicity of the in-plane components of the DM interactions, is shown by black line.}
\label{fig.str1}
\end{figure}

 Ba$_2$CuGe$_2$O$_7$ crystallizes in the tetrahedral structure (the space group is $P\overline{4}2_{1}m$, No.~113). The building blocks of Ba$_2$CuGe$_2$O$_7$ are the distorted CuO$_4$ tetrahedra, which are interconnected by the GeO$_4$ tetrahedra, as explained in Fig.~\ref{fig.str1}. The experimental lattice parameters are $a=8.466$ and $c=5.445$~\AA. Other parameters of the crystal structure can be found Ref.~\cite{exp_structure}.
\noindent

\par Most part of electronic structure calculations have been performed using plane-wave \texttt{Quantum ESPRESSO} (QE) method with ustrasoft pseudopotentials~\cite{QuantumE}. Some test calculations have been also performed using the full-potential linearized augmented-plane-wave method, as implemented in the \texttt{WIEN2k} package~\cite{Wien2k}, and the linear muffin-tin orbital (\texttt{LMTO}) method~\cite{LMTO1,LMTO2}. We employ the Perdew-Burke-Ernzerhof exchange-correlation functional within GGA~\cite{PBE} (except \texttt{LMTO}, where we use the Vosko-Wilk-Nusair functional~\cite{VWN}). All calculations have been performed on the mesh of $10 \times 10 \times 10$ ${\bf k}$-points in the Brillouin zone and the kinetic energy cutoff in the QE calculations is set to 90 Ry.

\par The electronic band structure obtained in the QE method with the SO coupling is shown in Fig.~\ref{fig.bands} (more accurate \texttt{WIEN2k} method provides essentially the same picture, as discussed in Supplementary Materials~\cite{SM}).
\noindent
\begin{figure}[t]
\begin{center}
\includegraphics[width=0.47\textwidth]{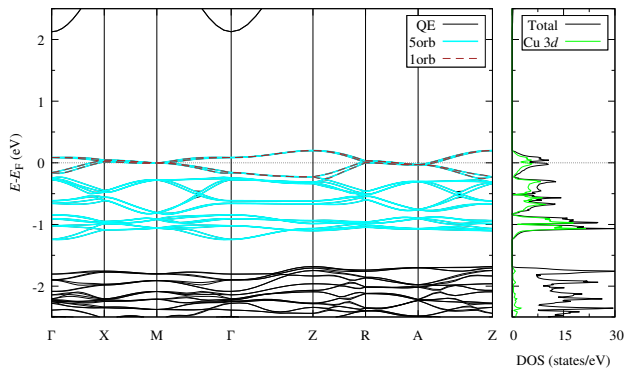}
\end{center}
\caption{
(Left panel) Electronic band structure of Ba$_2$CuGe$_2$O$_7$ with the spin-orbit coupling calculated within QE method as well as in the Wannier basis for the 1- and 5-orbital models. (Right panel) Total and partial Cu-$3d$ densities of states as obtained in the QE method. The Fermi level ($E_{\rm F}$) is at zero energy. Notations of the high-symmetry points of the Brillouin zone are taken from Ref.~\cite{BradlayCracknell}.}
\label{fig.bands}
\end{figure}
\noindent The electronic structure near the Fermi level features 20 bands (per two formula units) of predominantly Cu-$3d$ character, which are well isolated from other bands. These Cu-$3d$ bands can be used for the construction of the more general 5-orbital model (referring to the total number of $3d$ orbitals per one Cu site without spin). Furthermore, the Cu-$3d$ bands are split into two groups consisting of the 8 Cu-$e_{g}$ and 12 Cu-$t_{2g}$ bands, which are separated by the direct gap at around $-$$0.8$ eV. Finally, the Cu-$t_{2g}$ bands are further split due to the tetragonal distortion so that in the proximity of the Fermi level there are 4 half-filled Cu-$xy$ bands (which in the $xy$ coordinate frame have a large weight of the $x^2$-$y^2$ states). These bands are nearly separated from other Cu-$t_{2g}$ bands (by only slightly touching them around $\mathrm{Z}$ point of the Brillouin zone) and can be used as the basis for the construction of the minimal 1-orbital model.

\par The effective Hubbard-type model for these magnetic bands,
\noindent
\begin{widetext}
\begin{equation}
\hat{\cal{H}}  =  \sum_{ij} \sum_{\sigma \sigma'} \sum_{ab}
t_{ij}^{ab\sigma \sigma'}
\hat{c}^\dagger_{i a \sigma}
\hat{c}^{\phantom{\dagger}}_{j b \sigma'} +
  \frac{1}{2}
\sum_{i}  \sum_{\sigma \sigma'} \sum_{abcd} U^{abcd}
\hat{c}^\dagger_{i a \sigma} \hat{c}^\dagger_{i c \sigma'}
\hat{c}^{\phantom{\dagger}}_{i b \sigma}
\hat{c}^{\phantom{\dagger}}_{i d \sigma'},
\label{eqn.ManyBodyH}
\end{equation}
\end{widetext}
\noindent is formulated in the basis of Wannier functions constructed from the Bloch waves for either all 20 Cu-$3d$ bands or 4 Cu-$xy$ bands, which we will call the ``target bands''~\cite{review2008,WannierRevModPhys}. The operator $\hat{c}^\dagger_{i a \sigma}$ ($\hat{c}_{i a \sigma}$) in Eq.~(\ref{eqn.ManyBodyH}) stands for the creation (annihilation) of an electron with the spin $\sigma$ in the Wannier orbital $a$ of the site $i$. The noninteracting one-electron part of the model, $\hat{t}_{ij} = [ t_{ij}^{ab\sigma \sigma'} ]$, is associated with the matrix elements of the Kohn-Sham Hamiltonian in the Wannier basis. Since the latter is complete in the subspace of target bands, these bands are perfectly reproduced by the parameters $\hat{t}_{ij}$, as illustrated in Fig.~\ref{fig.bands}. The parameters of screened on-site Coulomb interactions, $\hat{U} = [U^{abcd}]$, where evaluated in the framework of constrained random-phase approximation (cRPA)~\cite{cRPA}.

\par Nowadays, the method of maximally localized Wannier functions~\cite{MarzariVanderbilt} is widely used in all kind of applications, which can be formulation in a small basis set of atomic or similar to them Wannier orbitals~\cite{WannierRevModPhys}. Therefore, as a first trial, we have employed for our purposes the standard MLWF technique~\cite{WannierRevModPhys}, as implemented in the \texttt{wannier90} package~\cite{wannier90}. This procedure is based on the minimization of the spread functional
\noindent
\begin{equation}
\Omega = \sum_{n} \left\langle ({\bf r} - {\bf r}_{0n})^2 \right\rangle_{n}
\label{eqn.spread}
\end{equation}
\noindent with the additional condition ${\bf r}_{0n} = \langle {\bf r} \rangle_{n} \equiv \bar{\bf r}_{n}$, which results in
\noindent
\begin{equation}
\Omega = \sum_{n} \left[ \langle r^2 \rangle -  \bar{\bf r}_{n}^2 \right].
\label{eqn.spread1}
\end{equation}
\noindent $\langle \dots \rangle_{n}$ in Eqs.~(\ref{eqn.spread}) and (\ref{eqn.spread1}) denotes the expectation value in the Wannier state $n \equiv (i a \sigma)$ and $\bar{\bf r}_{n}$ is the Wannier center. The main obstacle with the use of the maximally localized Wannier functions for the model (\ref{eqn.ManyBodyH}) is that they (and, therefore, the model Hamiltonian) do not necessary obey the symmetry of the considered system. This is the well-known problem of the MLWF calculations, which was encountered in many applications~\cite{SouzaMarzariVanderbilt,Sakuma}. We have found that the situation is particularly bad for the non-centrosymmetric non-polar compounds like Ba$_2$CuGe$_2$O$_7$, where in the process of minimization of $\Omega$, the Wannier centers are significantly shifted relative to the atomic positions, thus completely destroying the $P\overline{4}2_{1}m$ symmetry. The intuitive reason for it can be seen from the form of Eq.~(\ref{eqn.spread1}), where the additional shift of $\bar{\bf r}_{n}$ will minimize $\Omega$. Furthermore, the second term in Eq.~(\ref{eqn.spread1}) is not invariant under unitary transformation of the Wannier functions belonging to the same atomic site, which is clearly at odds with the fundamental requirement of rotational invariance of the model (\ref{eqn.ManyBodyH})~\cite{RInv}. Several solutions to circumvent this problem have been proposed in the literature, including symmetry-adapted MLWF~\cite{Sakuma} and selectively localized Wannier functions (SLWF)~\cite{SLWF}.

\par In this work, we propose a simple but very efficient procedure, where instead of treating all ${\bf r}_{0n}$ in Eq.~(\ref{eqn.spread}) as independent variables, we request them to be equal for each atomic site: ${\bf r}_{0n} \equiv {\bf r}_{0i}$. Note that, in our case, the Wannier functions serve as the basis of the model Hamiltonian (\ref{eqn.ManyBodyH}), where the individual positions $\bar{\bf r}_{n}$ are less important as they do not explicitly enter the construction of the model. More important is the subspace formed by the Wannier functions, which should be of the right symmetry. Therefore, to certain extent, it is wiser to reduce the number of variational parameters (which will inevitably lead to the increase of $\Omega$) for the sake of keeping the right symmetry of the model.

\par Then, the vector ${\bf r}_{0i}$ can be either a fixed input parameter (for instance, the position of the site $i$) or obtained variationally to minimize $\Omega$: $\partial \Omega / \partial {\bf r}_{0i} =0$, which naturally leads to the requirement ${\bf r}_{0i} = \frac{1}{N_{i}} \sum_{n \in i} \bar{\bf r}_{n}$, where $N_{i}$ is the number of the Wannier functions at the site $i$. In this case, $\Omega$ is still given by Eq.~(\ref{eqn.spread1}), but with ${\bf r}_{0i}$ instead of $\bar{\bf r}_{n}$'s. For the non-polar Ba$_2$CuGe$_2$O$_7$, the so obtained ${\bf r}_{0i}$ exactly coincides with the position $\boldsymbol{R}_{i}$ of the site $i$. Moreover, $\bar{\bf r}_{n} = {\bf r}_{0i}$ for all one-dimensional representations of the point group (formed by the $xy$, $x^2$-$y^2$, and $3z^2$-$r^2$ Wannier orbitals) and only for the two-dimensional representation (formed by the $yz$ and $zx$ orbitals), $\bar{\bf r}_{n}$'s are split around ${\bf r}_{0i}$ along the $z$ axis, as required by the symmetry. The obtained $\Omega$ is only slightly larger in comparison with the results of maximal localization procedure ($\Omega = 56.3$~\AA$^2$ for $N=20$ Cu-$3d$ functions, including spin, in comparison with $54.7$~\AA$^2$ in the MLWF method). However, the Wannier functions and the Hamiltonian (\ref{eqn.ManyBodyH}) obey the $P\overline{4}2_{1}m$ symmetry of the system, which is important improvement in comparison with the standard MLWF calculations.

\par For comparison, in the MLWF method, ${\bf r}_{0i}$'s also coincide with $\boldsymbol{R}_{i}$. Nevertheless, the individual Wannier centers $\bar{\bf r}_{n}$ are shifted away from $\boldsymbol{R}_{i}$, even for the one-dimensional representations, thus fully destroying the point-group symmetry. We have also applied the SLWF method by fixing all $\bar{\bf r}_{n}$ in $\boldsymbol{R}_{i}$ by the constraint condition. As expected, such requirement restores the point-group symmetry and also produces reasonable parameters of the spin model. Yet, we believe that our procedure is more natural and straightforward for these purposes as it does not introduce any extra steps, such as the constraint. Furthermore, the SLWF method yields slightly larger $\Omega = 56.5$~\AA$^2$.

\par Further details of implementation as well as the comparison with the MLWF and SLWF techniques are given in Supplementary Materials~\cite{SM}.

\

\section{\label{sec:symmetry} Form and symmetry properties of exchange interactions and electric polarization}
\subsection{\label{sec:general} General remarks}
\par
\begin{figure}[b]
\begin{center}
\includegraphics[width=0.47\textwidth]{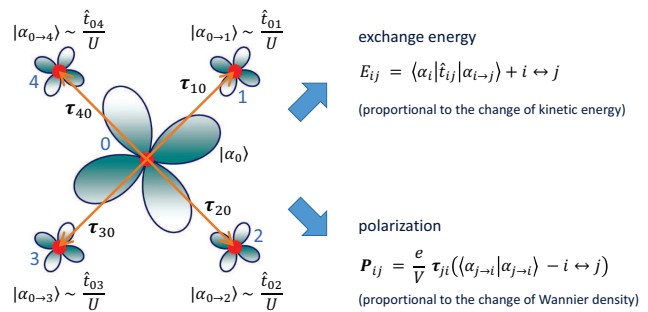}
\end{center}
\caption{
Schematic view on the superexchange theory for exchange interactions \textit{and} electric polarization: in the atomic limit, the hole is localized in the Wannier state $| \alpha_{0} \rangle$ of the central site $0$. Then, in the 1st order of perturbation theory with respect to the transfer integrals, $\hat{t}_{0j}$, this Wannier function acquires tails $| \alpha_{0 \to j} \rangle$ spreading to the neighboring sites $j$. By knowing the wavefunctions to the 1st order in $\hat{t}_{0j}$, one can evaluate the energy to the 2nd order in $\hat{t}_{0j}$, which constitutes the basis of the superexchange theory of the exchange interactions. Equivalently, the electric polarization can be related to the change of the Wannier density to the 2nd order in $\hat{t}_{0j}$. $\boldsymbol{\tau}_{ji}$ denotes the vector connecting the atomic site $i$ with the site $j$.}
\label{fig.SE}
\end{figure}

 The analytical expression relating the electric polarization with spin magnetization is a matter of controversy~\cite{KNB,Mostovoy,CuFeO2_Arima,MnI2_Xiang,Cu2OSeO3,PRB2014,PRB2017}. We believe that the most logical approach, at least for the localized electron systems, is the SE theory, which treats all transfer integrals as a perturbation in the 1st order of $\hat{t}_{ij}/U$~\cite{Anderson}. Here, we sketch the main ideas of this approach. All technical details can be found in Refs.~\cite{PRB2014,PRB2017,PRB2019,PRB2020}. The situation is schematically illustrated in Fig.~\ref{fig.SE}, where for simplicity we consider only average Coulomb repulsion $U$~\cite{footnote1}. Nevertheless, in all numerical calculations we take into account all necessary ingredients, including crystal-field splitting, SO and Hund's rule interactions~\cite{Khaliullin2009,PRB2015b,PRB2019}. Since ferroelectricity is the intrinsic property of insulating systems, the SE approximation is justified, at least as the starting point for analysis of the spin dependence of $\boldsymbol{P}$. Although the canonical SE theory deals with the energy~\cite{Anderson}, it can be naturally reformulated in terms of the Wannier functions $| w_{i} \rangle$ for the occupied states by considering the perturbation theory expansion for these Wannier functions.
\noindent

\par For the $d^9$ systems, like Ba$_2$CuGe$_2$O$_7$, the formulation is especially simple: in the atomic limit, the single hole resides in a Kramers doublet and is described by a pseudospin, so that by knowing the hole state $| \alpha_{i} \rangle$ at site $i$ one can automatically find the direction of the pseudospin at the same site as
\noindent
\begin{equation}
\boldsymbol{e}_{i} = \frac{\langle \alpha_{i} | \hat{\boldsymbol{\sigma}} | \alpha_{i} \rangle }{ |\langle \alpha_{i} | \hat{\boldsymbol{\sigma}} | \alpha_{i} \rangle | },
\label{eq:spine}
\end{equation}
\noindent where $\hat{\boldsymbol{\sigma}}$ is the vector of Pauli matrices. Hence, in the atomic limit, $| w_{i} \rangle = | \alpha_{i} \rangle$ is nothing but the occupied Wannier function associated with site $i$. Then, to the 1st order in $\hat{t}_{ij}$, $| w_{i} \rangle$ acquires tails spreading to the neighboring sites $j$  (see Fig.~\ref{fig.SE}):
\noindent
\begin{equation}
| w_{i} \rangle \approx | \alpha_{i} \rangle + \sum_{j} | \alpha_{i \to j} \rangle.
\label{eq:wtails}
\end{equation}
\noindent Furthermore, it is sufficient to consider only the transfer integrals connecting the occupied and unoccupied states: if both states are located in the occupied (or unoccupied) part, the corresponding contributions to the exchange interactions are cancelled out, being a general property of perturbation theory for the energy~\cite{Anderson}. Similar property holds for the electric polarization, which should remain invariant under unitary transformation of the occupied hole states $| \alpha_{i} \rangle$~\cite{FE_theory1,FE_theory2,FE_theory3}.

\par Then, knowing $| w_{i} \rangle$ to the 1st order in $\hat{t}_{ij}$, one can find the kinetic energy to the 2nd order in $\hat{t}_{ij}$ for any configuration of $\boldsymbol{e}_{i}$ and $\boldsymbol{e}_{j}$, which are specified by $| \alpha_{i} \rangle$ and $| \alpha_{j} \rangle$, respectively. This energy is further mapped onto the model
\noindent
\begin{equation}
\mathcal{H}^{\mathrm{S}} = \sum\limits_{\langle i j\rangle} \left( - J_{ij}\boldsymbol{e}_{i} \cdot \boldsymbol{e}_{j} + \boldsymbol{D}_{ij} \cdot [\boldsymbol{e}_{i}\times\boldsymbol{e}_{j}] + \boldsymbol{e}_{i} \cdot \tensor{\Gamma}_{ij} \boldsymbol{e}_{j} \right),
\label{eq:spinmodel}
\end{equation}
\noindent formulated in terms of the isotropic exchange constants $J_{ij}$, antisymmetric DM vectors $\boldsymbol{D}_{ij}$, and the traceless symmetric anisotropic tensors $\tensor{\Gamma}_{ij}$.

\par A good aspect of the SE theory is that, by using the same type of approximations for the Wannier function, one can also present $\boldsymbol{P}$ in the pairwise form, $\boldsymbol{P} = \sum_{\langle i j\rangle} \boldsymbol{P}_{ij}$, and derive an analytical expression for $\boldsymbol{P}_{ij}$, which is valid to the 2nd order in $\hat{t}_{ij}$. For these purposes, it is convenient to start with the ``Berry-phase expression'',
\noindent
\begin{equation}
\boldsymbol{P}=-\frac{e}{V}\sum\limits_{i}^{\mathrm{occ}} \langle w_{i} | {\bf r} | w_{i} \rangle
\label{eq:elpol}
\end{equation}
\noindent (where $-$$e$ is the electron charge and $V$ is the unit cell volume) and consider the asymmetric distribution of tails of the Wannier function in Eq.~(\ref{eq:wtails}), caused by the change of the magnetic order. Furthermore, one can write (relative to some central site $i$) ${\bf r} = \boldsymbol{\tau}_{ji} + \Delta {\bf r}_{j}$, where $\boldsymbol{\tau}_{ji} = \boldsymbol{R}_{j} - \boldsymbol{R}_{i}$, and assume that $| \Delta {\bf r}_{j} | \ll |\boldsymbol{\tau}_{ji}|$, which is equivalent to the statement that the weight of each Wannier function is distributed between lattice points located at $\boldsymbol{R}_{j}$. This is nothing but the standard requirement of discretization, inherent to the lattice model, which is widely used for the definition of the current operator and other physical quantities~\cite{Bari,Kubo}. For instance, the characteristic average radius of the Wannier function in Ba$_2$CuGe$_2$O$_7$ can be estimated as $\langle | \Delta {\bf r}_{j} | \rangle \sim \sqrt{\Omega / N} = 1.68$~\AA, while $|\boldsymbol{\tau}_{ji}|$ for the nearest neighbors in the tetragonal plane is about $6$~\AA. Under this assumption, the intersite matrix elements, $\langle \alpha_{i} | {\bf r} | \alpha_{i \to j} \rangle \approx \boldsymbol{\tau}_{ji} \langle \alpha_{i} | \alpha_{i \to j} \rangle$, vanish because of the orthogonality condition. The matrix elements $\langle \alpha_{i \to j} | {\bf r} | \alpha_{i \to j} \rangle \approx \boldsymbol{\tau}_{ji} \langle \alpha_{i \to j} | \alpha_{i \to j} \rangle$ are proportional to the Wannier density at the site $j$ and parallel to the bond. Since $\boldsymbol{\tau}_{ij} = - \boldsymbol{\tau}_{ji}$, the tail $| \alpha_{j \to i} \rangle$ will contribute to $\boldsymbol{P}_{ij}$ with the opposite sign (see Fig.~\ref{fig.SE}). This is different from the exchange energy, where the processes $j \to i$ and $i \to j$ are additive. Particularly, this means that the exchange interactions can be obtained already in the simplest 1-orbital model, while the polarization vanishes in the 1-orbital case and it is essential to work with the multiorbital models. Altogether, this leads to the following expression for the electric polarization~\cite{PRB2019,PRB2020}:
\noindent
\begin{equation}
\boldsymbol{P} = \sum\limits_{\langle i j\rangle} \boldsymbol{\epsilon}_{ji} \left( P_{ij} \boldsymbol{e}_{i} \cdot \boldsymbol{e}_{j} + \boldsymbol{\mathcal{P}}_{ij} \cdot [ \boldsymbol{e}_{i}\times\boldsymbol{e}_{j} ] + \boldsymbol{e}_{i} \cdot \tensor{\Pi}_{ij} \boldsymbol{e}_{j} \right),
\label{eq:spinpol}
\end{equation}
\noindent where $\boldsymbol{\epsilon}_{ji} = \boldsymbol{\tau}_{ji} / | \boldsymbol{\tau}_{ji} |$ is the unit vector along the bond $i$-$j$, $P_{ij}$ is the scalar, $\boldsymbol{\mathcal{P}}_{ij}$ is the vector, and $\tensor{\Pi}_{ij}$ is the traceless tensor, which appear, respectively, in the 0th, 1st, and 2nd order of the SO coupling. Thus, there is a direct analogy with the form of isotropic (Heisenberg-like), antisymmetric (DM-like) and anisotropic exchange interactions in Eq.~(\ref{eq:spinmodel}). However, unlike $\boldsymbol{D}_{ij}$, which is the antisymmetric vector with respect to the permutation of $i$ and $j$, $\boldsymbol{\mathcal{P}}_{ij}$ is the symmetric one due to the additional prefactor $\boldsymbol{\epsilon}_{ji}$ in Eq.~(\ref{eq:spinpol})~\cite{PRB2017,PRB2019}. Similarly, $P_{ij}$ is the antisymmetric scalar and $\tensor{\Pi}_{ij}$ is the antisymmetric tensor.

\par Eq.~(\ref{eq:spinpol}) has a clear physical meaning: each bond $i$-$j$ can be viewed as an electric dipole, in which the redistribution of charge between the poles $i$ and $j$ depends on the relative directions of spins. Therefore, it is natural that the direction of polarization in each dipole is parallel to the bond $i$-$j$. Furthermore, in such interpretation, the solid of ``electric dipoles'' does not necessarily imply the ``charge order'': since each pole of the dipole participates in several bonds (dipoles), the excess of the charge at certain atomic site in some bond can be compensated by its deficiency in another bond, being in line with the general definition of the electric polarization in terms of the macroscopic current flowing through the sample~\cite{FE_theory1,FE_theory2,FE_theory3}. The formula (\ref{eq:spinpol}) does not explicitly include the contributions of the oxygen (and other non-magnetic) sites, which seems to be at odds with phenomenological theories of the electric polarization based on the inverse DM~\cite{SergienkoPRB} and spin-current mechanism~\cite{KNB}. This is of course an approximation. However, absolutely the same level of approximations is used for derivation of the SE interactions in Eq.~(\ref{eq:spinmodel}). Therefore, if the model for the SE interactions is regarded to be acceptable, the same is expected for the model (\ref{eq:spinpol}) for the electric polarization. Below, we will show that the ME properties of Ba$_2$CuGe$_2$O$_7$ can be indeed described by Eq.~(\ref{eq:spinpol}). In Sec.~\ref{sec:other}, we will briefly discuss other mechanisms and contributions to $\boldsymbol{P}$, which are not included to Eq.~(\ref{eq:spinpol}), and try to resolve some controversies between our SE model and phenomenological theories~\cite{SergienkoPRB,KNB,Mostovoy}.

\subsection{\label{sec:polsi} Nonexistence of the single-site polarization for the spin $1/2$}
\par Eq.~(\ref{eq:spinpol}) takes into account only intersite contributions to the electric polarization, which depend on the relative orientation of spins in the bonds. Should it also include the single-site contributions, depending only on the directions of individual spins? Considering numerous attempts to interpret the ME properties of the Cu$^{2+}$ based spin-$1/2$ compounds in terms of such single-site effects~\cite{MurakawaPRB,Cu2OSeO3,Seki2012,YWLee,JTZhang,YNii}, the issue is indeed very controversial and the answer to this question is of principal importance.

\par As is well known, the single-site contribution to the exchange energy vanishes for the spin $1/2$, being one of fundamental consequences of Kramers degeneracy for systems with half-integer total spin~\cite{Kramers}. Now we will prove that a similar property holds for the single-site part of the polarization. The latter can be also derived from the general ``Berry-phase formula'' (\ref{eq:elpol}) and is given by $\boldsymbol{P} = -\frac{e}{V}{\rm Tr} \{ \hat{{\bf r}} \hat{\mathcal{D}} \}$, where $\hat{{\bf r}}$ is the matrix of the position operator in the basis of Kramers' states $| + \rangle$ and $| - \rangle$, forming the doublet and constructed from the Wannier functions at the given site, and $\hat{\mathcal{D}}$ is the density matrix for the hole state $| \alpha \rangle$ in the same basis~\cite{PRB2015}. This is a rigorous ``Berry-phase'' analog of phenomenological term for the electric polarization associated with the change of the metal-ligand $d$-$p$ hybridization due to the SO coupling~\cite{MurakawaPRB,PRB2015,CuFeO2_Arima}. Then, the hole state $| \alpha \rangle$ is a linear combination of $| + \rangle$ and $| - \rangle$, which also specifies the direction of spin $\boldsymbol{e}$ via Eq.~(\ref{eq:spine}). Hence, $\hat{\mathcal{D}}$ depends on $\boldsymbol{e}$ through the SU(2) rotation matrices, describing the transformation of $| \alpha \rangle$. Since the Kramers states are degenerate, the energy does not depend on $\boldsymbol{e}$ and there is no single-ion anisotropy term. Similar property holds for the position operator. Indeed, since $| + \rangle$ and $| - \rangle$ are related to each other by the transformations $\hat{T} | + \rangle = - | - \rangle$ and $\hat{T} | - \rangle = | + \rangle$, where $\hat{T} = i \hat{\sigma}_y \hat{K}$ is the time-reversal operation in terms of the spin Pauli matrix $\hat{\sigma}_y$ and the complex conjugation operator $\hat{K}$, we will have the following properties for any real spinless operator ${\bf r}$: $\langle + | {\bf r} | + \rangle = \langle - | {\bf r} | - \rangle$ and $\langle + | {\bf r} | - \rangle = 0$~\cite{footnote3}. Therefore, $\hat{{\bf r}}$ is proportional to the unity matrix, meaning that the single-site part of $\boldsymbol{P}$ does not depend on $\boldsymbol{e}$. This concludes our proof, which is one of the central results of our work.

\par The bond-dependent symmetric anisotropic tensors $\tensor{\Pi}_{ij}$ are formally allowed by the symmetry, but are of the 2nd order in the SO coupling and can be neglected for the purposes of our work~\cite{PRB2019,PRB2020}.

\subsection{\label{sec:ex} Exchange interactions}
\par
\begin{figure}[b]
\begin{center}
\includegraphics[width=0.47\textwidth]{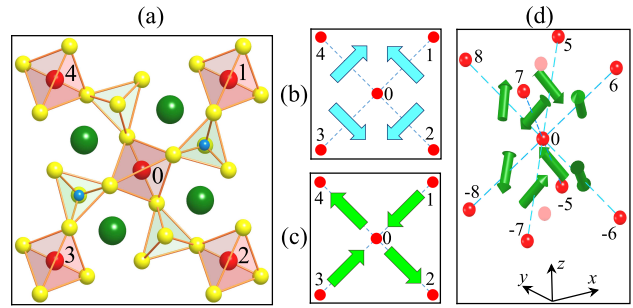}
\end{center}
\caption{
(a) Fragment of the crystal structure of Ba$_2$CuGe$_2$O$_7$ in the tetragonal plane, explaining the environment of the Cu sites $0$-$4$. (b) The in-plane components of Dzyaloshniskii-Moriya interactions $\boldsymbol{D}_{0j}$ operating in the nearest-neighbor Cu-Cu bonds $0$-$j$ (the transfer integrals $\boldsymbol{t}_{0j}$ in the 1-orbital model obey the same symmetry rules). (c) The vectors $\boldsymbol{\mathcal{P}}$ describing the antisymmetric part of electric polarization in the same bonds. (d) The symmetry properties of vectors $\boldsymbol{\mathcal{P}}$ in the next-nearest-neighbor bonds between the plane.}
\label{fig.str2}
\end{figure}

 The non-polar space group $P\overline{4}2_1m$ includes 8 symmetry elements, which can be obtained by combining 4 rotoinversion operations about $z$, $\hat{S}_{4}^{z}$, with $\hat{C}_{2}^{x}+ \left(\frac{1}{2},\frac{1}{2},0 \right)$ (the 2-fold rotation about $x$, followed by the shift in the units of lattice parameter $a$). They impose a symmetry constraint on the exchange interactions in Eq.~(\ref{eq:spinmodel}). Particularly, the DM interactions between nearest neighbors in the $xy$ plane are given by
\noindent
\begin{equation}
\boldsymbol{D}_{0j} = (-1)^{j} d^{xy} [\boldsymbol{\epsilon}_{j0} \times \boldsymbol{n}^{z}] + d^{z} \iota_{j0} \boldsymbol{n}^{z}
\label{eq:DM}
\end{equation}
\noindent (see Fig.~\ref{fig.str2} for the notations), where $\boldsymbol{n}^{z}$ is the unit vector along the $z$-axis, so that $[\boldsymbol{\epsilon}_{j0} \times \boldsymbol{n}^{z}]$ describes the regular 4-fold rotations in the $xy$ plane and the additional prefactor $(-1)^{j}$ arises from the rotoinversion symmetry transformation of the axial vectors $\boldsymbol{D}_{0j}$, and $\iota_{ij} = \frac{i-j}{|i-j|}$ is the antisymmetric scalar satisfying the conditions $\iota_{ij} = - \iota_{ji}$ and $|\iota_{ij}| = 1$. The prefactor $(-1)^{j}$ is responsible for a number of interesting effects, including possible formation of antiskyrmion textures~\cite{Bogdanov,Nayak,Huang}.
\noindent

\subsection{\label{sec:pol} Electric polarization}
\par Similar symmetry analysis can be performed for the polarization. Yet, the main difference between the exchange interactions and parameters $P_{ij}$, $\boldsymbol{\mathcal{P}}_{ij}$, and $\tensor{\Pi}_{ij}$ of the electric polarization is the symmetry properties with respect to the permutation of the atomic indices, which arise from the additional prefactor $\boldsymbol{\epsilon}_{ji}$ in Eq.~(\ref{eq:spinpol}): if the exchange interaction is symmetric, the corresponding to it parameter of the electric polarization should be antisymmetric, and vice versa. Therefore, the bond-dependence of these two groups of parameters will be generally different.

\par For the nearest-neighbor (nn) bonds in and between the tetragonal $xy$ planes, $P_{ij}$ vanishes due to the symmetry constraints imposed by the $P\overline{4}2_1m$ space group. We have found that the only sizable isotropic contributions come from the next-nn bonds $\left( \pm a/2, \pm a/2, \pm c \right)$ between the planes (see Fig.~\ref{fig.str2} for the notations of atomic sites). They are given by
\noindent
\begin{equation}
P_{ij} = (-1)^{j} \iota_{j0} p^{0}_{\perp}.
\label{eq:polnnnI}
\end{equation}

\par As for the antisymmetric part of $\boldsymbol{P}$ in the $xy$ plane, the nn contributions are allowed by the symmetry. The corresponding parameters $\boldsymbol{\mathcal{P}}_{ij}$ are given by
\noindent
\begin{equation}
\boldsymbol{\mathcal{P}}_{0j} = (-1)^{j} p^{xy}_{\parallel} \iota_{j0} \boldsymbol{\epsilon}_{j0}.
\label{eq:polnn}
\end{equation}
\noindent In addition to them, we have found sizable contributions operating in the next-nn bonds between the planes (see Fig.~\ref{fig.str2}d). Neglecting small difference between $x$- and $y$-components of $\boldsymbol{\mathcal{P}}_{ij}$, which are formally allowed by the symmetry, these parameters can be presented as
\noindent
\begin{equation}
\boldsymbol{\mathcal{P}}_{0j} \approx p^{xy}_{\perp} \iota_{j0} [\boldsymbol{\epsilon}_{j0} \times \boldsymbol{n}^{z}] + (-1)^{j} p^{z}_{\perp} \boldsymbol{n}^{z}.
\label{eq:polnnn}
\end{equation}

\section{\label{sec:model} Model analysis}
\subsection{\label{sec:1orbital} 1-orbital model}
\par In the 1-orbital case, we have two Wannier functions per each Cu site, which can be obtained by choosing the trial orbitals so that they correspond to the pure majority ($\uparrow$) and minority ($\downarrow$) spin states and then applying the regular procedure of the maximal localization~\cite{WannierRevModPhys}. This procedure should specify the phases of the Wannier functions and the Hamiltonian itself. The corresponding transfer integrals have the following form:
\noindent
\begin{equation}
\hat{t}_{ij} = t_{ij}^{0} \hat{\mathbb{1}} + i \boldsymbol{t}_{ij} \hat{\boldsymbol{\sigma}},
\label{eq:t1o}
\end{equation}
\noindent where $\hat{\mathbb{1}}$ is the $2$$\times$$2$ unit matrix and $\hat{\boldsymbol{\sigma}} = (\hat{\sigma}_{x},\hat{\sigma}_{y},\hat{\sigma}_{z})$ is the vector of Pauli matrices. To the lowest orders, $t_{ij}^{0}$ does not depends on the SO coupling, while $\boldsymbol{t}_{ij}$ emerges in the 1st order of it. Thus, in the 1-orbital model, the SO coupling contributes solely to the transfer integrals. With the proper choice of the phases, all the coefficients $t_{ij}^{0}$ and $\boldsymbol{t}_{ij} = (t_{ij}^{x},t_{ij}^{y},t_{ij}^{z})$ become real. In the following, we will call it a ``real representation''. Then, the Hermitian property of $\hat{t}_{ij}$ imposes the conditions: $t_{ji}^{0} = t_{ij}^{0}$ and $\boldsymbol{t}_{ji} = - \boldsymbol{t}_{ij}$. The vectors $\boldsymbol{t}_{ij}$ have the same form as $\boldsymbol{D}_{ij}$ (see Fig.~\ref{fig.str2}b) and for the nearest bonds in the $xy$ plane are given by
\noindent
\begin{equation}
\boldsymbol{t}_{0j} = (-1)^{j} t^{xy} [\boldsymbol{\epsilon}_{j0} \times \boldsymbol{n}^{z}] + t^{z} \iota_{j0} \boldsymbol{n}^{z}.
\label{eq:tvec1o}
\end{equation}
\noindent The corresponding parameters can be evaluated within QE method as: $t^{xy} = 5.4$, $t^{z} = -$$30.8$, and  $t^{0} = -$$37.0$ meV.

\par The electron-electron interactions in the 1-orbital model are specified by the single parameter $U$ of the on-site Coulomb repulsion between electrons with the opposite projections of spins, which can be evaluated within cRPA as $U \approx 3.74$ eV~\cite{cRPA}.

\par In the atomic limit, the single hole resides in the highest Kramers doublet, which is obtained after the diagonalization of the crystal field and the SO interaction, and the corresponding hole state $| \alpha_{i} \rangle$ specifies the direction of spin at site $i$, as was explained above. Then, one can formulate the SE theory by considering the virtual hoppings of electrons from all occupied states to $| \alpha_{i} \rangle$ (and back) as a perturbation~\cite{Anderson}. In the 1-orbital model this can be done analytically to obtain the following expressions for the exchange interactions~\cite{SM}:
\noindent
\begin{equation}
J_{ij} = -\frac{1}{U} \left( (t_{ij}^{0})^2 - \frac{1}{3} {\rm Tr} \left\{ \boldsymbol{t}_{ij} \otimes \boldsymbol{t}_{ij} \right\} \right),
\label{eq:J1o}
\end{equation}
\noindent
\begin{equation}
\boldsymbol{D}_{ij} = \frac{2t_{ij}^{0}}{U} \boldsymbol{t}_{ij},
\label{eq:D1o}
\end{equation}
\noindent and
\noindent
\begin{equation}
\tensor{\Gamma}_{ij} = \frac{2}{U} \left( \boldsymbol{t}_{ij} \otimes \boldsymbol{t}_{ij} - \frac{1}{3} {\rm Tr} \left\{ \boldsymbol{t}_{ij} \otimes \boldsymbol{t}_{ij} \right\} \tensor{\mathds{1}} \right),
\label{eq:G1o}
\end{equation}
with $\otimes$ denoting the direct product of two vectors and $\tensor{\mathds{1}}$ being the $3 \times 3$ unit tensor.

\par Since all exchange interactions in the bond are expressed in terms of (maximum) four parameters $(t_{ij}^{0},\boldsymbol{t}_{ij})$, they are not independent. Particularly, the tensor $\tensor{\Gamma}_{ij}$ is fully expressed in terms of the DM interactions and the isotropic exchange as~\cite{Shekhtman}:
\noindent
\begin{equation}
\tensor{\Gamma}_{ij} = -\frac{1}{2J_{ij}} \left( \boldsymbol{D}_{ij} \otimes \boldsymbol{D}_{ij} - \frac{1}{3} {\rm Tr} \left\{ \boldsymbol{D}_{ij} \otimes \boldsymbol{D}_{ij} \right\} \tensor{\mathds{1}} \right),
\label{eq:G1o2}
\end{equation}
\noindent which means that the 1-orbital model~(\ref{eq:t1o}) is subjected to hidden symmetries~\cite{Kaplan,Shekhtman} (see also Supplementary Materials~\cite{SM} for the discussion of how the independent parameters of the transfer integrals can be generally found irrespectively of the phases of the Wannier functions). Even more generally, one can argue that by means of unitary transformations (corresponding to rotations of the spin variables) the Hamiltonian~(\ref{eq:t1o}) \textit{in each separate bond} can be reduced to $\hat{\tilde{t}}_{ij} = \tilde{t}_{ij}^{0} \hat{\mathbb{1}}$~\cite{Kaplan}, so that the bond would be totally described by an isotropic exchange coupling only. However, whether this can be done \textit{simultaneously for all bond} depends on the symmetry of the system. In our case, the only possibility is to use different unitary transformations at two Cu sites in the unit cell: Cu1 (corresponding to $i=0$ in Fig.~\ref{fig.str2}) and Cu2 ($i=1$-$4$). In this case, one can eliminate $t^{z}$ (and therefore $d^{z}$), which is the same for all nn bonds $0$-$j$ [see Eq.~(\ref{eq:tvec1o})]. On the contrary, the contributions of $t^{xy}$ enter Eq.~(\ref{eq:tvec1o}) with different signs and cannot be eliminated. The corresponding unitary transformation is given by $\hat{U}_{S} = {\rm diag}(e^{\mp i \psi/2},e^{\pm i \psi/2})$, where the upper (lower) sign stands for the site Cu1 (Cu2) and $\psi = - \tan^{-1} \frac{t^{0}t^{z}}{(t^{0})^2-(t^{z})^2}$. Then, in the new coordinate frame we have $\tilde{t}^{0} = -\sqrt{(t^{0})^2 + (t^{z})^2}$, while $\tilde{t}^{z} = 0$. Furthermore, it is straightforward to see that the remaining $\boldsymbol{t}_{0j} = (-1)^{j} t^{xy} [\boldsymbol{\epsilon}_{j0} \times \boldsymbol{n}^{z}]$ is translationally invariant on the lattice with only one Cu site in unit cell (see Fig.~\ref{fig.str1}): for instance, the translation of the bond $3$-$0$ ($4$-$0$) to the bond $0$-$1$ ($0$-$2$) does not change $\boldsymbol{t}_{ij}$. The corresponding parameters of isotropic and DM nn interactions in the plane can be evaluated using Eqs.~(\ref{eq:J1o}) and (\ref{eq:D1o}) as $J = -$$0.616$ and $d^{xy}=0.140$ meV, respectively.

\par Considering only $J$ and $d^{xy}$, the classical magnetic ground state corresponds to a spiral with spins rotating in the plane, which can be specified by the vector $\boldsymbol{n}^{\perp} = (-\sin \phi, \cos \phi, 0)$ being perpendicular to the plane. The corresponding propagation vector can be easily found by considering the symmetry properties of $\boldsymbol{D}_{0j}$, which yield~\cite{SM}:
\noindent
\begin{equation}
\boldsymbol{q} = \boldsymbol{q}_{0} + \delta \boldsymbol{q},
\label{eq:propv}
\end{equation}
\noindent where $\boldsymbol{q}_{0} = (2\pi, 0, 0)$ (in the units of $1/a$) corresponds to the two-sublattice AFM order in the plane, while $\delta \boldsymbol{q} = (\delta q \sin \phi, \delta q \cos \phi, 0)$ with $\delta q = - \sqrt{2} d^{xy}/J$ describes its modulation caused by the spin spiral. Furthermore, the classical spin-spiral energy does not depend on $\phi$, which can be arbitrary~\cite{SM}. Nevertheless, it is very important that not only $\boldsymbol{n}^{\perp}$, but also $\delta \boldsymbol{q}$ depends on $\phi$. For instance, by varying $\phi$ one can change the type of the spin spiral from cycloidal ($\delta \boldsymbol{q} \perp \boldsymbol{n}^{\perp}$), realized for $\boldsymbol{n}^{\perp} = \frac{1}{\sqrt{2}}(\pm 1, \pm 1, 0)$, to proper screw ($\delta \boldsymbol{q} \parallel \boldsymbol{n}^{\perp}$), realized for $\boldsymbol{n}^{\perp} = (\pm 1, 0, 0)$ and $(0, \pm 1, 0)$. Such behavior is due to the rotoinversion symmetry: for instance, if instead of the rotoinversion we dealt with the regular 4-fold rotation around $z$, the spin spiral would be cycloidal for all $\phi$. This symmetry is also crucially important for the behavior of electric polarization, which will be considered in Sec.~\ref{sec:sppol}.

\par The nonvanishing matrix elements of the anisotropy tensor $\tensor{\Gamma}_{ij}$ satisfy the following properties: $\Gamma^{xx} = \Gamma^{yy} = -\frac{1}{2} \Gamma^{zz} = 0.003$ meV, which holds for all nn bonds, and $\Gamma^{xy}_{0j} = (-1)^{j} \Gamma^{xy}$ for $j= 1$-$4$ in Fig.~\ref{fig.str2}, where $\Gamma^{xy} = 0.008$ meV. The compass-type anisotropy, $\Gamma^{xy}$, does not contribute to the classical ground state energy. Then, positive $\Gamma^{xx} = \Gamma^{yy}$ in combination with the antiferromagnetic $J$ would lead to the easy-plane AFM configuration, which can be indeed stabilized by applying the external magnetic field along $z$~\cite{Zheludev1997,Chovan}. For the classical spins, the configuration remains degenerate with respect to the in-plane rotations of the magnetization. However, the zero-point motion, which is the first quantum correction to the classical ground state energy to the 1st order in $1/S$, lifts this degeneracy and stabilizes the magnetization parallel to one of the square diagonals in the $xy$ plane~\cite{Yildirim}. This corresponds to $\boldsymbol{n}^{\perp} = \frac{1}{\sqrt{2}}(\pm 1, \pm 1, 0)$. The quantitative estimates of this effect, following the work of Yildirim~\textit{et al.}~\cite{Yildirim}, can be found in Supplementary Materials~\cite{SM}. Similar strategy has been applied recently for the analysis of SO interaction driven magnetic properties of iridate Ba$_2$IrO$_4$~\cite{Katukuri}. Thus, in view of these arguments, the ground state is expected to be cycloidal with $\delta \boldsymbol{q} \perp \boldsymbol{n}^{\perp}$, in agreement with the experimental situation~\cite{Zheludev1998}. Due to the DM interaction, the spins at the neighboring sites along $\delta \boldsymbol{q}$ are additionally rotated relative to each other by the angle $\vartheta = |d^{xy}/J| \sim 13.0^\circ$, which is close to the experimental value of $9.7^\circ$~\cite{Zheludev1998}.

\par Furthermore, $\tensor{\Gamma}_{ij}$ is responsible for anharmonic modulations in the spiral structure. This purely classical effect, which is driven by $\Gamma^{zz}$, is described by sine-Gordon equations~\cite{Zheludev1998}, quantifying the preferential grouping of spins closer to the easy plane (also known as ``bunching''~\cite{REbunching1,REbunching2}). The quantitative analysis is given in Supplementary Materials~\cite{SM}: the anisotropy $\Gamma^{zz}$ indeed leads to the visible anharmonicity of the spin-spiral pattern, but has little effect on its periodicity.

\par In fact, the model (\ref{eq:spinmodel}) with the additional constraint (\ref{eq:G1o2}) has been intensively studied in the literature. Further details for Ba$_2$CuGe$_2$O$_7$ can be found in Ref.~\cite{Chovan}.

\par Thus, the simple 1-orbital model is very useful for unveiling basic magnetic properties of Ba$_2$CuGe$_2$O$_7$: in this case all exchange interactions can be obtained analytically, which allows an easy and transparent interpretation. However, from the viewpoint of quantitative analysis, the abilities of the 1-orbital model are quite limited, while more general 5-orbital model is believed to be more appropriate for these purposes. Even more importantly, the orbital degrees of freedom are indispensable for the magnetic part of $\boldsymbol{P}$: since the polarization is antisymmetric with the site indices in $\hat{t}_{ij}$ (see Fig.~\ref{fig.SE}), it vanishes in the 1-orbital case~\cite{PRB2014,PRB2019}. It can be paraphrased differently: because of this antisymmetry, $\boldsymbol{P}$ appears to be proportional to intra-atomic Hund's coupling $J_{\rm H}$ (in an analogy with compass-type exchange interactions in iridates~\cite{Khaliullin2009}), which is absent in the 1-orbital model. Therefore, in the next Section we turn to the analysis of the 5-orbital model.

\subsection{\label{sec:5orbital} 5-orbital model}
\par The one-electron part of the model Hamiltonian (\ref{eqn.ManyBodyH}) was constructed from the electronic structure obtained in the QE calculations with the SO coupling, which contributes to both site-diagonal and off-diagonal elements of $t_{ij}^{ab\sigma \sigma'}$. Other options are discussed in the Supplementary Materials~\cite{SM}. The crystal field splits the atomic $3d$ level in four groups located at $-$$0.36$, $-$$0.34$, $0.08$, and $0.55$ eV. The first two are the $e_{g}$ levels of the $3z^2$-$r^2$ and $x^2$-$y^2$ symmetry, which are followed by two $t_{2g}$ levels standing, respectively, for the degenerate $yz$/$zx$ and nondegenerate $xy$ states (also having large weight of the $x^2$-$y^2$ states in the $xy$ coordinate frame shown in Fig.~\ref{fig.str1}). The SO interaction constant is about $0.12$ eV, which is comparable with the splitting of the $t_{2g}$ levels.

\par The matrices of screened Coulomb interactions obtained within cRPA~\cite{cRPA} were fitted in terms of three independent parameters, specifying the interactions among $3d$ electrons in the spherical case~\cite{review2008}: the Coulomb repulsion $U=F^0 \approx 4.05$ eV, the intra-atomic (Hund's) exchange coupling $J_{\rm H} = (F^2 + F^4)/14 \approx 0.97$ eV, and the nonsphericity $B = (9F^2 - 5F^4)/441 \approx 0.1$ eV, where $F^0$, $F^2$, and $F^4$ are the screened radial Slater's integrals. Quite expectedly, the value of $U$ is larger than in the 1-orbital case due to the reduced number of channels available for the screening in cRPA. This screening is not particularly strong in the case of cuprates: the Cu $3d$ band is nearly filled, thus leaving only a small number of holes available for the screening, which explains relatively large values of $U$~\cite{PRB2018}. In order to fulfil the symmetry requirements of Ba$_2$CuGe$_2$O$_7$ in our SE calculations, we have uses the simplified form of $\hat{U} = [U^{aabb}]$, which was given by only $U$ and $J_{\rm H}$ as $U^{aabb} = U$ and $U^{abba} = J_{\rm H}$ (for $a \ne b$).

\par Then, by applying the SE theory for the exchange interactions~\cite{PRB2015b,PRB2019}, one obtains the following parameters of the spin model: $J = -0.430$, $d^{xy} = 0.109$, and $d^{z} = -0.007$ meV. The obtained $J$ exceeds the experimental value nearly by a factor of 2~\cite{Zheludev1999,footnote2}. Nevertheless, $\vartheta = |d^{xy}/J| \sim 14.5^\circ$ is consistent with the experimental value of $9.7^\circ$~\cite{Zheludev1998}, meaning that our $d^{xy}$ is also overestimated. Yet, we would like to emphasize that in comparison with the 1-orbital model, the Coulomb $U$ rises by only 8\%, while the AFM $J$ drops by more than 30\%. This means that, beside the AFM contribution to $J$ (being proportional to $1/U$~\cite{KugelKhomskii}), in the 5-orbital model there is also the ferromagnetic (FM) one ($\sim J_{\rm H}/U^2$), which substantially improves the quantitative description.

\par The DM interaction $d^{z}$, which is responsible for the spin canting and net magnetic moment in the $xy$ plane, is small and does not play a decisive role. The parameters of the exchange anisotropy tensor $\tensor{\Gamma}_{ij}$ are comparable with those of the 1-orbital model: $\Gamma^{xx} = \Gamma^{yy} = -\frac{1}{2} \Gamma^{zz} = 0.003$ meV and $\Gamma^{xy} = 0.005$ meV. Therefore, the anisotropic properties in the $xy$ plane as well as the anharmonicity of the spin-spiral pattern are expected to be similar to the ones for the 1-orbital model and we do not consider them here. Furthermore, there are small matrix elements $\Gamma_{0j}^{xz} = \Gamma_{0j}^{zx} = \sqrt{2}\epsilon_{j0}^{x}\Gamma^{xz}$ and $\Gamma_{0j}^{yz} = \Gamma_{0j}^{zy} = \sqrt{2}\epsilon_{j0}^{y}\Gamma^{xz}$, where $\Gamma^{xz} = 0.003$ meV. However, they do not play a major role.

\par The next important isotropic interaction after $J$ is that between the 2nd neighbors in the adjacent layers (or next-nn), $J_{\perp}^{2} = -0.010$ meV. This interaction is AFM and, in combination with $J$, stabilizes the \textit{ferromagnetic} alignment between the layers, in agreement with the experiment~\cite{Zheludev1998}. The coupling between the 1st neighbors is weakly ferromagnetic ($J_{\perp}^{1} \sim 10^{-4}$ meV). Thus, without SO coupling, the magnetic structure would be $C$-type AFM, in which the AFM spin ordering in the $xy$ plane coexists with the FM stacking along $z$. The corresponding N\'eel temperature can be estimated in the framework of random phase approximation~\cite{tyab,TCRPA} as $T_{\rm N} \approx 12$~K, which exceeds the experimental value of $3.2$~K~\cite{Zheludev1998}, probably due to the overestimation of $J$ and $J_{\perp}^{2}$. Similar problem was encountered for Ba$_2$CoGe$_2$O$_7$~\cite{PRB2015}. Formally, the quantitative description can be improved by decreasing the value of $U$ and thus increasing the FM contribution to $J$ via the change of the ratio $J_{\rm H}/U^2$~\cite{KugelKhomskii}. Similar tendency was found for the electric polarization, which will be discussed in Sec.~\ref{sec:sppol}.

\section{\label{sec:elpol} Magnetism and electric polarization}
\subsection{\label{sec:sppol} Spontaneous polarization induced by cycloidal order}
\par First, let us consider the behavior of electric polarization $P^{z}$ induced by the spin-spiral order~\cite{MurakawaPRL,MurakawaPRB}. According to Eq.~(\ref{eq:spinpol}), only those bonds, which are directed out of the $xy$ plane and have finite projection on $z$, can contribute to $P^{z}$. We have found that the main such contributions are associated with the next-nn bonds between adjacent tetragonal planes (see Fig.~\ref{fig.str2}d). Other contributions are either small or not effective: for instance, the atoms in the nn bonds $(0,0,\pm c)$ are always ferromagnetically coupled and these bonds do not contribute to the magnetic dependence of $\boldsymbol{P}$.

\par In fact, $P^{z}$ is a combining effect: $P^{z} = P^{z}_{A} + P^{z}_{I}$, where both contributions are induced by the spiral spin order. The first one is driven by the antisymmetric ($A$) mechanism, which is described by Eq.~(\ref{eq:polnnn}), in combination with the main Eq.~(\ref{eq:spinpol}). This is an analog of the DM interaction for the exchange energy, which can be related to the intrinsic spin current flowing in the system~\cite{Kikuchi}. Then, considering the ideal spin spiral, specified by the rotation plane with $\boldsymbol{n}^{\perp} = (-\sin \phi, \cos \phi, 0)$ and the propagation vector (\ref{eq:propv}), it is straightforward to find that~\cite{SM}
\noindent
\begin{equation}
P^{z}_{A} = - \frac{4 \sqrt{2} ac}{a^2+2c^2} \frac{d^{xy}}{J} p^{xy}_{\perp} \sin 2 \phi.
\label{eq:PzA}
\end{equation}
\noindent We would like to emphasize here that the $\sin 2 \phi$ dependence of $P^{z}_{A}$ is the combination of two, multiplying each other, factors~\cite{SM}: quite naturally, $[\boldsymbol{e}_{i} \times \boldsymbol{e}_{j}]$ depends on $\phi$, specifying the orientation of the spin-rotation plane, and alone would result in the regular $\sin \phi$ (or $\cos \phi$) dependence of $P^{z}_{A}$ (apart from a phase). However, in addition to that, the spin-spiral propagation vector $\boldsymbol{q}$ also depends on $\phi$ via $\sin \phi$ and $\cos \phi$, being the result of the rotoinversion symmetry (see discussions in Sec.~\ref{sec:1orbital}). Altogether, they yield $\sin 2 \phi$ in Eq.~(\ref{eq:PzA}).

\par The second mechanism is isotropic and described by Eq.~(\ref{eq:polnnnI}), again in combination with Eq.~(\ref{eq:spinpol}). The key point here is that the spin spiral breaks the tetragonal symmetry so that the angles between spins in the direction of propagation $\boldsymbol{q}$ and the perpendicular to it direction are different. Therefore, the cancellation of contributions coming from these two types of bonds does not occur, leading to finite total polarization, which can be evaluated as~\cite{SM}
\noindent
\begin{equation}
P^{z}_{I} = - \frac{2 \sqrt{2}c}{\sqrt{a^2+2c^2}} \left( \frac{d^{xy}}{J} \right)^2 p^{0}_{\perp} \sin 2 \phi.
\label{eq:PzI}
\end{equation}
\noindent One can find some analogy with the skyrmion compounds, like GaV$_4$S$_8$, where the DM interactions give rise to either cycloidal or skyrmion order. These magnetic orders are manifested in the change of $\boldsymbol{P}$ originating from the competition of antisymmetric and isotropic mechanisms~\cite{PRB2019,PRB2020}. The main difference is that GaV$_4$S$_8$ is a polar compound, where the spontaneous electric polarization emerges below certain structural transition point and is further modulated by the magnetic order, while in Ba$_2$CuGe$_2$O$_7$ the polarization is solely induced by the spin-spiral order.

\par
\begin{figure}[t]
\begin{center}
\includegraphics[width=0.47\textwidth]{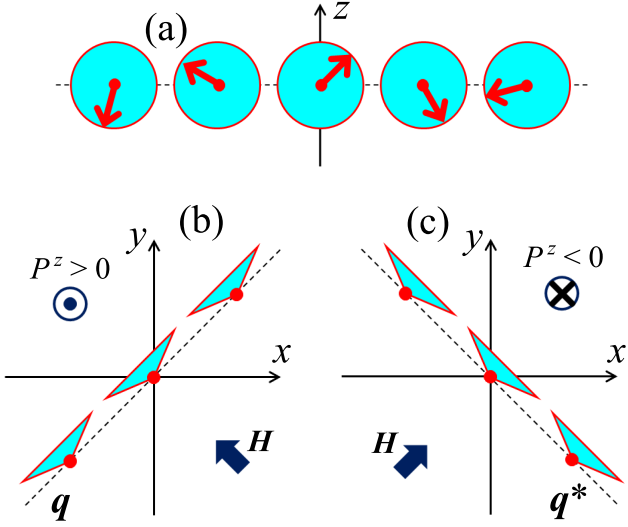}
\end{center}
\caption{
(a) Side view and (b,c) top views on cycloids with the propagation vectors $\boldsymbol{q}$ (b) and $\boldsymbol{q}^{*}$ (c). By applying the magnetic field perpendicular to the spin rotation plane, one can switch between magnetic domains with $\boldsymbol{q}$ and $\boldsymbol{q}^{*}$, and thus reverse the electric polarization $P^{z}$.}
\label{fig.Pz}
\end{figure}

 The $\sin 2 \phi$ dependence of both $P^{z}_{A}$ and $P^{z}_{I}$ nicely reproduces the experimental behavior of Ba$_2$CuGe$_2$O$_7$~\cite{MurakawaPRL}. Namely, rotating the helical spin plane by the magnetic field, one can switch the direction of propagation of the spin spiral from $\boldsymbol{q} = (2 \pi + \frac{\delta q}{\sqrt{2}}, \frac{\delta q}{\sqrt{2}}, 0)$ to $\boldsymbol{q}^{*} = (2 \pi - \frac{\delta q}{\sqrt{2}}, \frac{\delta q}{\sqrt{2}}, 0)$, which leads to the reversal of $P^{z}$ (see Fig.~\ref{fig.Pz}). Alternatively, one can control the direction of $P^{z}$ by applying the external electric field and thus switch the direction of propagation of spins between $\boldsymbol{q}$ and $\boldsymbol{q}^{*}$~\cite{MurakawaPRL}.
\noindent

\par Another important question is whether the proposed SE theory is able to reproduce the experimental value of $|P^{z}| \sim 0.3$ $\mu$C/m$^2$~\cite{MurakawaPRL} and what is the relative strength of $P^{z}_{A}$ and $P^{z}_{I}$? Using the numerical values of the parameters, one can obtain the following estimates (at $\phi = \pi/4$): $P^{z}_{A} \sim 0.505 \, p^{xy}_{\perp}$ and $P^{z}_{I} \sim -0.086 \, p^{0}_{\perp}$. The values of the parameters $p^{xy}_{\perp}$ and $p^{0}_{\perp}$ appear to be sensitive to the details of calculations. The upper estimate was found in the LMTO method yielding $p^{xy}_{\perp} = 0.217$ and $p^{0}_{\perp} = -0.036$ $\mu$C/m$^2$~\cite{SM}, which correspond to $P^{z} = 0.11$  $\mu$C/m$^2$, being in reasonable agreement with the experimental value. The QE estimate is considerably lower: $p^{xy}_{\perp} = 0.046$ and $p^{z}_{\perp} = 0.037$ $\mu$C/m$^2$~\cite{SM}, corresponding to $P^{z} = 0.02$  $\mu$C/m$^2$, which is an order of magnitude smaller than the experimental value.

\par This comparison may be viewed as somewhat discouraging, especially because QE is typically regarded as more reliable method in comparison with LMTO~\cite{SM}. Nevertheless, one should keep in mind that, at least formally, this discrepancy can be easily cured by considering rather modest correction of the model parameters. For instance, in the first approximation, $p^{0}_{\perp}$ is proportional to $J_{\rm H}/U^3$~\cite{PRB2019}. Similar behavior is expected for $p^{xy}_{\perp}$. While $J_{\rm H}$ is typically well defined (and close to the atomic value), $U$ is frequently treated as an adjustable parameter on the semi-empirical level~\cite{RInv}. Then, if we wanted to reproduce the experimental value of $P^{z}$ in the QE method simply by adjusting the value of $U$, we would have to decrease it by factor $10^{\frac{1}{3}} \approx 2$ (i.e., to about 2 eV), which is quite a normal practice in the electronic structure calculations, for instance the ones based on the local density approximation $+$$U$ method~\cite{RInv}. Of course, there may be other factors, hampering the agreement with the experimental data, including validity of the SE approximation and necessity to consider the effects of higher orders in the $\hat{t}_{ij}/U$ expansion~\cite{PRB2015b}. Furthermore, there are lattice effects~\cite{SergienkoPRB,Malashevich}, which can dominate over the electronic ones considered in the present study.

\par In any case, the inequality $P^{z}_{A} \gg |P^{z}_{I}|$ means that the electric polarization induced by the spiral order in Ba$_2$CuGe$_2$O$_7$ stems from the antisymmetric spin-current mechanism. This naturally explains the difference between Ba$_2$CoGe$_2$O$_7$ and Ba$_2$CuGe$_2$O$_7$. In the former case, the polarization is substantially larger as it is driven by the single-site mechanism, which is permitted for the spin-$3/2$ and typically dominates in comparison with the intersite contributions~\cite{MurakawaPRL2,PRB2015}. In the spin-$1/2$ compound Ba$_2$CuGe$_2$O$_7$, the single-site term vanishes resulting in the drop of the polarization. Nevertheless, it is important that even in Ba$_2$CuGe$_2$O$_7$ the experimental polarization remains finite, thus indicating that there is another mechanism of the ME coupling besides the single-site one. Similar arguments apply for the magnetic ground state of Ba$_2$CuGe$_2$O$_7$ and Ba$_2$CoGe$_2$O$_7$: the latter is $C$-type antiferromagnet in spite of DM interactions operating in this compound similar to Ba$_2$CuGe$_2$O$_7$. However, the effect of DM interactions is suppressed by large single-site anisotropy, driving this system into the commensurate $C$-type AFM state. In Ba$_2$CuGe$_2$O$_7$, the single-site anisotropy is absent and the DM interactions start to dominate. Therefore, the regular ground state will be the incommensurate spin spiral, while the $C$-type AFM order is stabilized only in the external magnetic field. The behavior of electric polarization, accompanying this transition, will be considered in the next section.

\subsection{\label{sec:meeffect} Reorientation of polarization associated with the incommensurate-commensurate transition}
\par
\begin{figure}[b]
\begin{center}
\includegraphics[width=0.47\textwidth]{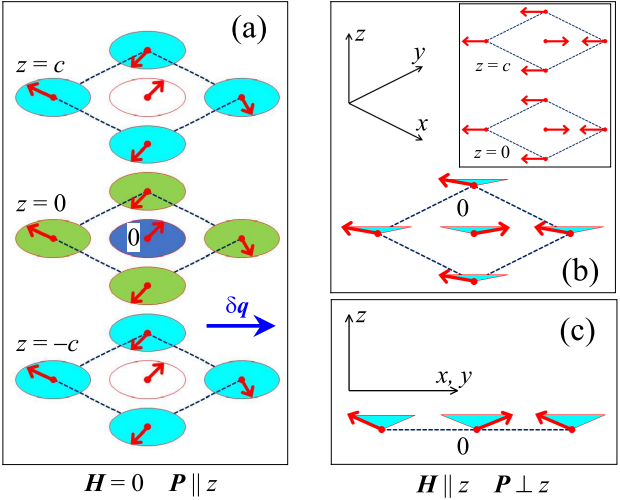}
\end{center}
\caption{
Summary of incommensurate-commensurate transition in Ba$_2$CuGe$_2$O$_7$ induced by the magnetic field $\boldsymbol{H} \parallel z$: the incommensurate cycloidal phase (a) is transformed into the $C$-type antiferromagnetic phase, which is illustrated in the inset of (b). The magnetic field leads to the canting of spins as explained in (b) and (c). The transition is accompanied by the flip of electric polarization from $\boldsymbol{P} \parallel z$ to $\boldsymbol{P} \perp z$, which is driven the spin-current mechanism. The magnetic sites surrounding the central atom $0$, which contribute to the electric polarization in the cycloidal and $C$-type antiferromagnetic phases are denoted by the cyan and green colors, respectively. $\delta \boldsymbol{q}$ specifies the direction of propagation vector.}
\label{fig.icc}
\end{figure}

 The application of the magnetic field $H$ along the $z$ axis in Ba$_2$CuGe$_2$O$_7$ gives rise to the incommensurate-commensurate (IC-C) transition from cycloidal to $C$-type AFM phase~\cite{Zheludev1997} (Fig.~\ref{fig.icc}). In the $C$ phase, the magnetic moments lie in the $xy$ plane, as requested by the exchange anisotropy, while the magnetic field leads to the small FM canting of spins along $z$. The latter is specified by the component $e^{z}$ of the spin direction vector, which is proportional to $H$ and can be found from the equilibrium condition as $e^{z} = - \frac{\mu_{\rm B}H}{8J}$.
\noindent

\par Below, we will argue that the IC-C transition should be accompanied by the reorientation of the polarization from the $z$ direction ($P^{z}$) into the $xy$ plane ($\boldsymbol{P}^{xy}$), which was overlooked in previous studies. These two components of the electric polarizations, $P^{z}$ and $\boldsymbol{P}^{xy}$, have rather different origins and are associated with different bonds. As we have seen in the previous section, the polarization $P^{z}$ is the multiferroic feature, which is induced solely by the cycloidal order without external field. On the contrary, $\boldsymbol{P}^{xy}$ is the manifestation of more conventional ME effect in Ba$_2$CuGe$_2$O$_7$, which is induced by the magnetic field $H$ and proportional to this field. In this case, $H$ not only stabilizes the $C$-type AFM state, but also breaks the symmetry of this state, similar to the conventional ME effect~\cite{DzyaloshinskiiME}. Indeed, the in-plane orientation of spins lowers the point-group symmetry of Ba$_2$CuGe$_2$O$_7$ from $\hat{S}_{4}^{z}$ (the 4-fold rotoinversion axis) to $\hat{T}\hat{C}_{2}^{z}$ (time reversal times 2-fold rotation about $z$). Then, the FM canting of spins along $z$ further breaks the $\hat{T}\hat{C}_{2}^{z}$ symmetry, thus allowing for the electric polarization in the $xy$ plane.

\par
\begin{figure}[t]
\begin{center}
\includegraphics[width=0.3\textwidth]{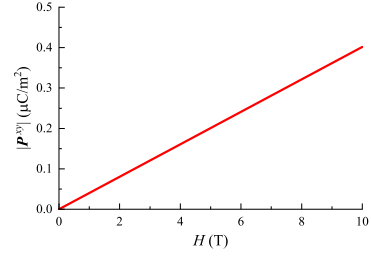}
\end{center}
\caption{
Magnetoelectric effect in Ba$_2$CuGe$_2$O$_7$: absolute value of electric polarization in the $xy$-plane versus magnetic field.}
\label{fig.meH}
\end{figure}

\begin{figure}[t]
\begin{center}
\includegraphics[width=0.47\textwidth]{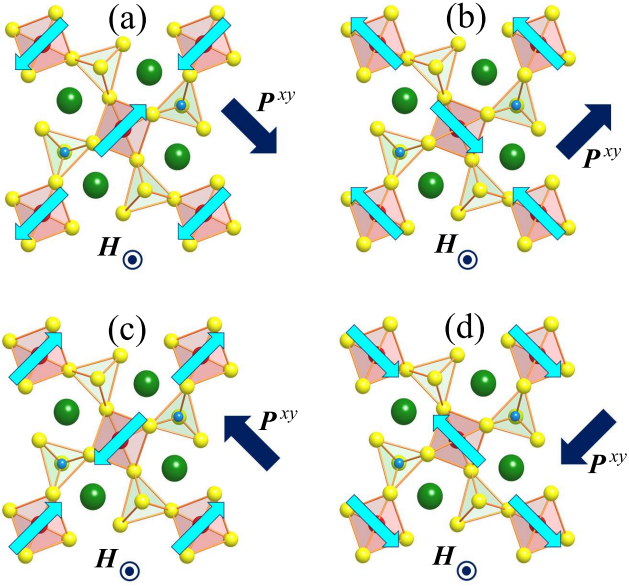}
\end{center}
\caption{
Magnetoelectric effect in Ba$_2$CuGe$_2$O$_7$: directions of electric polarization in the $xy$-plane, $\boldsymbol{P}^{xy}$, corresponding to different types of antiferromagnetic domains for the same direction of the external magnetic field $\boldsymbol{H} = (0,0,H)$. The directions of spin moments are shown by cyan arrows.}
\label{fig.me}
\end{figure}

 If $P^{z}$ originates from the next-nn Cu-Cu bonds, connecting adjacent $xy$ planes and therefore having finite $z$ component, the main contribution to $\boldsymbol{P}^{xy}$ is associated with the nn bonds in the plane (Fig.~\ref{fig.icc}). Assuming that, to the 1st order in $e^{z} \sim H$, the direction of spin at the central site is $\boldsymbol{e}_{0} = (\cos \phi, \sin \phi, e^{z})$ and the one at the neighboring sites is $\boldsymbol{e}_{j} = (-\cos \phi, -\sin \phi, e^{z})$, $\boldsymbol{P}^{xy}$ can be easily evaluated using Eq.~(\ref{eq:spinpol}) as~\cite{SM}:
\noindent
\begin{equation}
\boldsymbol{P}^{xy} = -\frac{\mu_{\rm B}Hp^{xy}_{\parallel}}{2J} (\cos \phi, -\sin \phi, 0),
\label{eq:Pxy}
\end{equation}
\noindent where $p^{xy}_{\parallel}$ is estimated within the QE method as $0.597$ $\mu$C/m$^2$ (other methods provide rather consistent description and the results are summarized in the Supplementary Materials~\cite{SM}). The obtained dependence of $|\boldsymbol{P}^{xy}|$ on $H$ is shown in Fig.~\ref{fig.meH}.
\noindent
\noindent Then, in the ground state configuration, $\phi = \pi/4$ (modulo $\pi/2$)~\cite{SM}, $\boldsymbol{P}^{xy}$ is perpendicular to the directions of spins. The reversal of all spins in the $C$ state, $\boldsymbol{e} \rightarrow -\boldsymbol{e}$, also reverses the direction of the polarization $\boldsymbol{P}^{xy} \rightarrow -\boldsymbol{P}^{xy}$. Alternatively, by applying the electric field, one can change the direction of polarization $\boldsymbol{P}^{xy}$ and thus switch between different AFM domains as illustrated in Fig.~\ref{fig.me}.
\noindent
\noindent This provides the possibility to control the directions of the antiferromagnetically coupled moments by the electric field.

\subsection{\label{sec:other} Other mechanisms and controversies}
\par In this section, we briefly discuss other contributions to the magnetically induced polarization (\ref{eq:spinpol}) and try to resolve some controversies between the SE theory, that we propose, and phenomenological theories~\cite{SergienkoPRB,KNB,Mostovoy}, which are frequently used for the interpretation of the experimental data. There is a widespread believe that a noncollinear alignment of spins induces the polarization perpendicular to the bond~\cite{KimuraARMR,CheongMostovoy,Khomskii,SergienkoPRB,KNB,Mostovoy}. Certainly, this is very different from the conclusion of our SE theory, where the polarization is expected to be parallel to the bond. Apparently, we are dealing we different mechanisms and each of the theories, including ours, is still incomplete for describing the behavior of the electric polarization.

\par For instance, the inverse DM mechanism~\cite{SergienkoPRB} was proposed to explain the emergence of electric polarization in response to the cycloidal spin order -- the only experimental possibility, which was known at that time. Then, the intermediate oxygen atoms are expected to move perpendicular to the bonds to acquire the energy gain associated with the so-induced DM interactions. Hence, $\boldsymbol{P}$ should be perpendicular to the bond. Nevertheless, similar arguments apply for the proper-screw spiral, where spins rotate in the plane perpendicular to the bonds. Then, the oxygen atoms are expected to move parallel to the bonds, which should be also the new direction of $\boldsymbol{P}$.

\par The electronic mechanism by Katsura, Nagaosa, and Balatsky (KNB) is based on the analysis of simple cluster model consisting of two transition-metal sites and intermediate oxygen site~\cite{KNB}. In this case, one can apply the symmetry arguments, similar to the ones considered by Dzyaloshinskii for Cr$_2$O$_3$~\cite{DzyaloshinskiiME}, and argue that there should be both transversal and longitudinal ME effect. In fact, the relative strength of these two effects in Cr$_2$O$_3$ strongly depends on the temperature and magnetic field, controlling the spin-flop transition~\cite{MECr2O3}.

\par The key assumption of our work is $\langle | \Delta {\bf r}_{j} | \rangle \ll |\boldsymbol{\tau}_{ji}|$, which was necessary for the discretization and construction of the lattice model for $\boldsymbol{P}$ (see Sec.~\ref{sec:general}). Of course, this is an approximation and the ratio $\langle | \Delta {\bf r}_{j} | \rangle / |\boldsymbol{\tau}_{ji}|$ for the nearest neighbors in the plane of Ba$_2$CuGe$_2$O$_7$ is $0.28$ (i.e., quite large). Under this assumption we were able to get rid of the ${\bf r}$-dependence in all matrix elements, including the intersite ones, $\langle \alpha_{i} | {\bf r} | \alpha_{i \to j} \rangle$, which is relevant to the KNB mechanism, but vanishes in our case.

\par One can draw again an analogy with the exchange interactions in insulators: the SE theory accounts only for kinetic contributions to Eq.~(\ref{eq:spinmodel}). However, the exchange interactions are not limited by the SE processes and there is still an ongoing discussion on whether and how these SE interactions should be combined with other contributions, for instance - the direct exchange interactions, which can be responsible for the ferromagnetism~\cite{Ku,Danis,PRB2015c}. In our case, the FM direct exchange could indeed improve the agreement with experimental data for $J$ and other parameters of exchange interactions~\cite{PRB2015}. By knowing the Wannier functions at the transition-metal sites, one can evaluate the direct exchange integral~\cite{Ku,Danis}, which can be additionally screened by the oxygen band~\cite{PRB2015c}. Similar situation is realized here: Eq.~(\ref{eq:spinpol}) can be viewed as the magnetic part of the polarization induced by the SE processes, while $\langle \alpha_{i} | {\bf r} | \alpha_{i \to j} \rangle$ are the  direct contributions, associated with the overlap of the Wannier functions. Note that Wannier functions have tails, spreading to the oxygen and other intermediate sites, and these tails mainly contribute to $\langle \alpha_{i} | {\bf r} | \alpha_{i \to j} \rangle$. Formally, these contributions are of the order of $\langle | \Delta {\bf r}_{j} | \rangle / |\boldsymbol{\tau}_{ji}|$ and, from this point of view, can be regarded as small corrections to our SE theory. However, they appear in the 1st order of $\hat{t}_{ij}/U$ (while the SE contributions -- only in 2nd) and, therefore, can be large. Nevertheless, one should also keep in mind that, similar to the direct exchange interactions, the integrals $\langle \alpha_{i} | {\bf r} | \alpha_{i \to j} \rangle$ can be additionally screened by the oxygen band (as the polarization in this band will be also affected by the magnetic order), while the bare values of $\langle \alpha_{i} | {\bf r} | \alpha_{i \to j} \rangle$ are probably overestimates.

\par Below, we evaluate the change of $P^{z}$ caused by bare integrals $\langle \alpha_{i} | {\bf r} | \alpha_{i \to j} \rangle$, which were calculated in the 5-orbital model between neighboring sites in the tetragonal plane. Neglecting small symmetric anisotropic part, $P^{z}$ in this case is given by $P^{z} \approx \sum_{ \langle ij \rangle } \left( \mathsf{P}_{ij} \boldsymbol{e}_{i} \cdot \boldsymbol{e}_{j} + \boldsymbol{\mathbbm{P}}_{ij} \cdot [\boldsymbol{e}_{i} \times \boldsymbol{e}_{j}] \right)$, where $\mathsf{P}_{0j} = (-1)^{j} \mathsf{p}^{0}_{\parallel}$ and $\boldsymbol{\mathbbm{P}}_{0j} = \mathsf{p}^{xy}_{\parallel} [\boldsymbol{\epsilon}_{j0} \times \boldsymbol{n}^{z}] + (-1)^{j} \mathsf{p}^{z}_{\parallel} \iota_{j0} \boldsymbol{n}^{z}$. The parameters of this model can be estimated within the QE method as $\mathsf{p}^{0}_{\parallel} = 1.382$, $\mathsf{p}^{xy}_{\parallel} = 0.454$, and $\mathsf{p}^{z}_{\parallel} = 0.039$ $\mu$C/m$^2$. We note that this $\mathsf{p}^{xy}_{\parallel}$ is comparable with the parameter $p^{xy}_{\parallel}$ obtained in the SE approximation and describing the behavior of electric polarization in the same bonds. However, $\mathsf{p}^{xy}_{\parallel}$ and $p^{xy}_{\parallel}$ are responsible for completely different effects, and it is important that $\mathsf{p}^{xy}_{\parallel}$ can contribute to $P^{z}$, which was observed experimentally. Then, the spin-spiral order gives rise to the polarization $P^{z} = P^{z}_{I} + P^{z}_{A}$ (i.e., the isotropic and antisymmetric contributions, both induced \emph{perpendicular} to the bonds). The analytical expressions for $P^{z}_{I}$ and $P^{z}_{A}$ can be obtained along the same line as described in Supplementary Materials~\cite{SM}, which yields $P^{z}_{I} = - \left( d^{xy}/J \right)^{2} \mathsf{p}^{0}_{\parallel} \sin 2 \phi$ and $P^{z}_{A} = - 2 \left( d^{xy}/J \right) \mathsf{p}^{xy}_{\parallel} \sin 2 \phi$. The origin of $P^{z}_{A}$ is similar to the KNB theory, but obeying the symmetry properties of Ba$_2$CuGe$_2$O$_7$. The $\sin 2 \phi$ dependence of $P^{z}_{A}$ and $P^{z}_{I}$ is consistent with the experimental behavior for Ba$_2$CuGe$_2$O$_7$, similar to the SE contribution given by Eq.~(\ref{eq:PzA}).

\par Using bare $\mathsf{p}^{xy}_{\parallel}$, $P^{z}_{A}$ can be estimated (at $\phi = \pi/4$) as $0.23$ $\mu$C/m$^2$, which alone is consistent with the experimental value of $P^{z}$ and could improve the relatively poor agreement with the experiment in the QE method (see Sec.~\ref{sec:sppol}). Nevertheless, $P^{z}_{A}$ is not the only multiferroic effect originating from $\langle \alpha_{i} | {\bf r} | \alpha_{i \to j} \rangle$: in addition to $P^{z}_{A}$, there is also isotropic contribution $P^{z}_{I}$, which obeys the same symmetry properties. If $P^{z}_{A}$ was anticipated by the KNB theory, $P^{z}_{I}$ was not considered before in any model of electric polarization induced by the magnetic order. $P^{z}_{I}$ has the opposite sign and can be estimated (again at $\phi = \pi/4$ and using bare $\mathsf{p}^{0}_{\parallel}$) as $-0.09$ $\mu$C/m$^2$. Thus, there will be a strong cancellation of isotropic and antisymmetric contributions, leading again to relatively small total value of $P^{z} = 0.14$ $\mu$C/m$^2$. Furthermore, we believe that bare $\mathsf{p}^{0}_{\parallel}$ and $\mathsf{p}^{xy}_{\parallel}$ should be additionally screened by the oxygen band, which we do not consider in our model.

\par Thus, the superexchange, as any model approach, does not necessarily include all possible contributions to the magnetic state dependence of $\boldsymbol{P}$. The main advantage of the SE theory, that we propose~\cite{PRB2014,PRB2017,PRB2019}, over other existing models~\cite{CheongMostovoy,Khomskii,TokuraSekiNagaosa,SergienkoPRB,KNB,Mostovoy} is that (i) it allows us to treat isotropic, antisymmetric, and symmetric anisotropic contributions on an equal footing; (ii) all these contributions obey the symmetry rules and, thus, can be applied for the semi-quantitative analysis of properties of the real materials; and (iii) all contributions can be easily evaluated using the same set of parameters as for the interatomic SE interactions. Nevertheless, this is basically a toy-model, which is helpful for understanding the behavior of $\boldsymbol{P}$, but does not substitute brute-force numerical calculations based on the Berry-phase formalism~\cite{FE_theory1,FE_theory2,FE_theory3}, where all the contributions and ingredients are automatically taken into account.

\section{\label{sec:summary} Conclusions}
\par We have considered general principles for constructing the spin models for the electric polarization in solids, which can be applied for the analysis of magnetoelectric coupling in the wide class of multiferroic compounds. One of crucial findings of our study is nonexistence of single-site contributions to the magnetic dependence of $\boldsymbol{P}$ for the spin $1/2$. This is an analog of the well-known theorem for the magnetic energy, which allows us to rule out the theories, where $\boldsymbol{P}$ at a certain magnetic site is solely determined by the direction of magnetization at the same site, from the interpretation of properties of multiferroic materials hosting one unpaired electron or hole in their magnetic building block. The principle should apply for all kind of lattices of Cu$^{2+}$, Ni$^{3+}$, V$^{4+}$, and Ti$^{3+}$ ions~\cite{MurakawaPRB,Cu2OSeO3,Seki2012,YWLee,JTZhang,YNii}, as well as the molecular complexes like the ($M_4$S$_4$)$^{5+}$ clusters in the lacunar spinel compounds Ga$M_4$S$_4$ ($M=$ V or Mo), which attracted a great deal of attention due ability to control the electric polarization by manipulating the skyrmionic texture~\cite{PRB2019,PRB2020,gavs2}. Another major breakthrough is the SE theory of electric polarization that we propose~\cite{PRB2019,PRB2020}. Contrary to phenomenological theories~\cite{SergienkoPRB,KNB,Mostovoy}, this is the first systematic strategy for deriving microscopic models aiming to describe spin dependencies of the $\boldsymbol{P}$. Over the years, the SE theory had enormous success in various applications for the analysis of interatomic magnetic interactions in insulators~\cite{Anderson,Moriya_weakF,Khaliullin2009} and, thus, well suits for the description of magnetically induced ferroelectricity -- yet another property, which is inherent to insulating compounds. The theory has been successfully applied for the analysis of electric polarization induced by complex magnetic orders in Ba$_2$CuGe$_2$O$_7$. We have argued that this and similar spin-$1/2$ materials can be used as testbed systems for exploration of spin-current driven ME phenomena as other mechanisms in them are either weak or forbidden by the symmetry. Particularly, the cycloidal order in Ba$_2$CuGe$_2$O$_7$ yields spontaneous electric polarization along the crystallographic $z$ axis, which can be inverted by rotating the propagation vector $\boldsymbol{q}$ within the tetragonal plane~\cite{MurakawaPRL}. In addition to that, we have predicted the flip of the polarization into the $xy$-plane in the course of the incommensurate-commensurate (cycloidal-AFM) transition in the magnetic field, where the direction of polarization in the plane can be further controlled by rotating the antiferromagnetically coupled spins in the same plane. Moreover, the origin of in-plane and out-off-plane polarizations is ultimately related to the spin-current mechanism operating in two groups of magnetic bonds situated, respectively, in and between the planes. Finally, we have proposed simple but efficient procedure for calculating the Wannier functions with proper point-group symmetry, which is crucially important for applications of this technique for the construction and analysis of microscopic models on the basis of first-principles electronic structure calculations.

\section*{Acknowledgement}
\par We are grateful to Professor Peter Kr\"uger for valuable discussions and careful reading of our manuscript. I.S. was supported by program AAAA-A18-118020190095-4 (Quantum).


\begin{thebibliography}{99}
\bibitem{footnote4}
Note that in metals the polarization is completely screened by free electrons.
\bibitem{footnote1}
$U = \frac{1}{M^2} \sum_{a=1}^{M} \sum_{b=1}^{M} U^{aabb}$, where $M=N_{i}/2$ is the number of orbitals at the atomic site $i$ without spin.
\bibitem{footnote3} The state of the Kramers doublet can be presented as $|\pm \rangle \equiv |\pm $$\uparrow \rangle \oplus |\pm$$\downarrow \rangle$, explicitely referring two parts of the vector with the spins $\uparrow$ and $\downarrow$, which leads to the identity $\langle + | {\bf r} | - \rangle =  \langle +$$\downarrow$$| {\bf r} | -$$\downarrow \rangle + \langle +$$\uparrow$$| {\bf r} | -$$\uparrow \rangle$. Then, $| - \rangle = -\hat{T} | + \rangle$ yields $| -$$\uparrow \rangle = -\hat{K}| +$$\downarrow \rangle$ and $| -$$\downarrow \rangle = \hat{K}| +$$\uparrow \rangle$, and therefore $\langle + | {\bf r} | - \rangle =  \langle +$$\downarrow$$| {\bf r} \hat{K} | +$$\uparrow \rangle - \langle +$$\uparrow$$| {\bf r} \hat{K} | +$$\downarrow  \rangle$, which is zero.
\bibitem{footnote2}
To be consistent with our definition, Eq.~(\ref{eq:spinmodel}), the experimental $J$ should be additionally multiplied by factor $-S^2 = -1/4$, which yields $J = -0.24$ meV.
\end{thebibliography}
\end{document}